\documentclass[adp,a4paper,fleqn%
]{w-art}
\usepackage{times,cite,w-thm}
\theoremstyle{plain}

\theoremstyle{definition}

\usepackage[]{graphicx}
\usepackage{dcolumn}
\usepackage{bm}
\usepackage{psfrag}
\usepackage{subfigure}

\newcommand{\be}{\begin{eqnarray}}
\newcommand{\ee}{\end{eqnarray}}
\newcommand{\bn}{\begin{eqnarray*}}
\newcommand{\en}{\end{eqnarray*}}

\newcommand{\nn}{\nonumber \\}
\newcommand{\nl}{\\}

\renewcommand{\vec}[1]{\mbox{\boldmath$#1$}}
\renewcommand{\d}{\mbox{\rm d}}

\renewcommand{\th}{\ensuremath{\theta}}

\newcommand{\ph}{\ensuremath{\phi}}

\newcommand{\al}{\ensuremath{\alpha}}
\newcommand{\bt}{\ensuremath{\beta}}
\newcommand{\sg}{\ensuremath{\sigma}}
\newcommand{\gm}{\ensuremath{\gamma}}
\newcommand{\dl}{\ensuremath{\delta}}
\newcommand{\lm}{\ensuremath{\lambda}}
\newcommand{\om}{\ensuremath{\omega}}
\newcommand{\Lm}{\ensuremath{\Lambda}}

\newcommand{\Sg}{\ensuremath{\Sigma}}
\newcommand{\Gm}{\ensuremath{\Gamma}}
\newcommand{\Om}{\ensuremath{\Omega}}

\newcommand{\ze}{\ensuremath{\hat{0}}}

\newcommand{\pvec}{\ensuremath{\vec{p}}}
\newcommand{\Pvec}{\ensuremath{\vec{P}}}

\newcommand{\Rvec}{\ensuremath{\vec{R}}}

\newcommand{\alvec}{\ensuremath{\vec{\al}}}
\newcommand{\sgvec}{\ensuremath{\vec{\sg}}}

\newcommand{\Avec}{\ensuremath{\vec{A}}}
\newcommand{\Dvec}{\ensuremath{\vec{D}}}

\newcommand{\Xvec}{\ensuremath{\vec{X}}}
\newcommand{\Wvec}{\ensuremath{\vec{W}}}
\newcommand{\Gmvec}{\ensuremath{\vec{\Gm}}}
\newcommand{\Omvec}{\ensuremath{\vec{\Omega}}}
\newcommand{\Sgvec}{\ensuremath{\vec{\Sg}}}

\newcommand{\nabvec}{\ensuremath{\vec{\nabla}}}

\newcommand{\hb}{\ensuremath{\hbar}}
\newcommand{\ihb}{\ensuremath{i \hbar}}
\newcommand{\pt}[1]{\ensuremath{{\partial \over \partial #1}}}

\newcommand{\lt}{\ensuremath{\left}}
\newcommand{\rt}{\ensuremath{\right}}

\renewcommand{\d}{\mbox{\rm d}}

\begin{document}
\DOIsuffix{theDOIsuffix}
\Volume{16}
\Month{01}
\Year{2007}
\pagespan{1}{}
\Receiveddate{XXXX}
\Reviseddate{XXXX}
\Accepteddate{XXXX}
\Dateposted{XXXX}
\keywords{Casimir invariance breakdown, elementary particles, curved space-time, non-inertial effects.}



\title[Breakdown of Casimir Invariance in Curved Space-Time]
{Breakdown of Casimir Invariance in Curved Space-Time}


\author[D. Singh]{Dinesh Singh\inst{1,}%
  \footnote{Corresponding author\quad E-mail:~\textsf{dinesh.singh@uregina.ca},
            Phone: +001\,306\,585\,4681,
            Fax: +001\,306\,585\,5659}}
\address[\inst{1}]{Department of Physics, University of Regina \\
Regina, Saskatchewan, S4S 0A2, Canada}
\author[N. Mobed]{Nader Mobed\inst{1}%
  \footnote{\quad E-mail:~\textsf{nader.mobed@uregina.ca},
            Phone: +001\,306\,585\,4359,
            Fax: +001\,306\,585\,5659}}


\begin{abstract}
It is shown that the commonly accepted definition for the Casimir scalar operators of the
Poincar\'{e} group does not satisfy the properties of Casimir invariance when applied to the
non-inertial motion of particles while in the presence of external
gravitational and electromagnetic fields, where general curvilinear co-ordinates are used to
describe the momentum generators within a Fermi normal co-ordinate framework.
Specific expressions of the Casimir scalar properties are presented.
While the Casimir scalar for linear momentum remains Lorentz invariant in the
absence of external fields, this is no longer true for the spin Casimir scalar.
Potential implications are considered for the propagation of photons, gravitons, and gravitinos
as described by the spin-3/2 Rarita-Schwinger vector-spinor field.
In particular, it is shown that non-inertial motion introduces a frame-based effective mass
to the spin interaction, with interesting physical consequences that are explored in detail.
\end{abstract}
\maketitle                   






\section{Introduction}

In a recent paper 
\cite{Singh-Mobed}, a question is raised about the limitations of the Poincar\'{e}
group for identifying the dynamics of a quantum mechanical particle in non-inertial motion.
Specifically, this paper considers the viability of applying the Pauli-Lubanski spin vector to describe the intrinsic
properties of a spin-1/2 particle while propagating along a general worldline in Minkowski space-time.
By expressing the Poincar\'{e} group generators in terms of curvilinear co-ordinates, to best accommodate
the symmetries of the particle's motion with respect to some fixed laboratory frame, it is shown that
the associated Casimir scalar invariant for particle spin becomes {\em frame-dependent}.
This is due to an additive term involving the Pauli spin vector $\sgvec$ coupled to a Hermitian three-vector
$\Rvec$ called the {\em non-inertial dipole operator} \cite{Singh-Mobed,Singh1,Singh2},
which relates the interaction between the momentum operators $\Pvec$ described by curvilinear co-ordinates.
Dimensionally, $|\Rvec| \sim |\Pvec|/r$, where $r$ is the local radius of curvature defined with respect
to the laboratory frame.
In the limit as $r \rightarrow \infty$, it follows that $\Rvec \rightarrow \vec{0}$ for fixed momentum,
and for a particle in rectilinear motion whose trajectory is best represented by Cartesian co-ordinates, $\Rvec = \vec{0}$
identically.
In this context, ``non-inertial motion'' refers exclusively to motion involving local rotational acceleration
of the frame comoving with the particle, since linear accelerated motion as described in terms of Rindler co-ordinates,
for example, naturally implies that $\Rvec = \vec{0}$, as the corresponding momentum generators automatically commute.

Similar considerations about Casimir operators defined in curved space-time and/or for accelerated
motion already exist in the literature.
One perspective \cite{Low} explores the properties of projective representations of the inhomogeneous
Hamilton group---which describe non-inertial frames according to classical Hamiltonian mechanics
and whose subgroup is the Galilei group for inertial frames---as applied to a quantum non-inertial state.
This examination finds that the Casimir invariants derived according to the Hamilton group lead to a
generalized description of spin as a Galilei invariant that is applicable to both inertial and non-inertial states.
A second perspective \cite{Gronwald,Lemke} makes the point that, within the context of gauge-field approaches to gravitation,
Casimir invariance of the Poincar\'{e} group is violated for a metric-affine gravity theory,
which must then be sought after via the more general symmetry group $GL(4,R)$.

The purpose of this paper is to explore the so-called Casimir properties of the Poincar\'{e} group for elementary particles,
while propagating non-inertially in the presence of external gravitational and electromagnetic fields.
A detailed first-principles computation of the spin-1/2 Casimir scalars for linear momentum and
spin angular momentum in the presence of gravitational and electromagnetic fields is presented in Sec.~\ref{Sec.Casimir.spin-1/2},
which includes the formalism of Fermi normal co-ordinates, along with its generalization of
curvilinear spatial co-ordinates \cite{Singh-Mobed}.
A demonstration of the explicit breakdown of Casimir invariance for the relevant operators concerning linear momentum and
spin angular momentum in the Poincar\'{e} group due to non-inertial motion in the absence of external fields
is then given in Sec.~\ref{Sec:Casimir-violation}.
This is followed by Sec.~\ref{Sec.Casimir.higher-spin}, which outlines the principles of computing the
Casimir scalars for massive particles of higher-order spin with explicit computations presented for spin-1 to spin-2 inclusive,
where the Rarita-Schwinger vector-spinor field \cite{Rarita} is assumed for determining the spin-3/2 case.
The diagonalized form of the Casimir spin scalar operators, in the limit of vanishing external fields, is then presented
in Sec.~\ref{Sec.Casimir.diagonlized-spin}, with an exploration of interesting physical consequences for both massive and
massless particles, where the latter set describes the predicted behaviour of spin-1 photons, spin-3/2 gravitinos predicted in
supersymmetry, and spin-2 gravitons while in non-inertial motion.
A discussion of more general considerations motivated by the main results in this paper is presented in Sec.~\ref{Sec.discussion},
followed by a brief conclusion in Sec.~\ref{Sec.conclusion}.

\section{Casimir Scalars of Spin-1/2 Particles in Curved Space-Time}
\label{Sec.Casimir.spin-1/2}

\subsection{Fermi Normal Co-ordinates}

Before exploring the Casimir properties of spin-1/2 particles in curved space-time following the approach taken in this paper,
it is useful to first briefly review the formalism of Fermi normal co-ordinates as commonly presented \cite{Poisson},
along with its generalization to incorporate curvilinear spatial co-ordinates \cite{Singh-Mobed2}.

\subsubsection{Formalism}

The location and the orientation of the frame locally comoving with the spin-1/2 particle
is determined using the orthonormal tetrad formalism \cite{Singh-Mobed}, in which general curvilinear co-ordinates
identify the particle frame's spatial location in flat space-time with respect to the stationary laboratory frame.
It follows that the spatial co-ordinates associated with the comoving frame are naturally chosen to be either Cartesian for rectilinear motion,
or some other set, such as cylindrical or spherical co-ordinates, if it better reflects the symmetry of the particle's motion between
neighbouring intervals of proper time.

A similar approach can be taken when generalizing the problem to incorporate space-time curvature \cite{Singh-Mobed2},
where the spatial part of the comoving frame is described by Fermi normal co-ordinates $X^\mu = \lt(T, \Xvec\rt)$,
and the Riemann curvature tensor is projected onto the associated worldline.
While it is possible to find exact expressions of Fermi normal co-ordinates for certain curved backgrounds \cite{Chicone},
generally they are represented as perturbative expansions away from the origin identified by the four-velocity of
the reference worldline for a freely falling observer \cite{Manasse}.
Generalizations that include both curvature and inertial contributions also exist \cite{Ni,Li,Huang}, in which the latter
are due to non-geodesic motion.
In particular, it is shown \cite{Li,Huang} that explicit coupling of gravitational and inertial effects appear for third-order
expansions of the metric with respect to $\Xvec$.
While it is useful to consider both types of contributions for this problem at a later date, for the sake of clarity
the inertial effects are neglected here.
In what follows, geometric units of $G = c = 1$ are assumed throughout, where the Riemann and Ricci tensors are defined according to the
conventions of Misner, Thorne, and Wheeler \cite{MTW}, but with a metric signature of $-2$.
Space-time co-ordinates are labelled from $0$ to $3$ by Greek indices, where $\bar{\mu}$ denotes general co-ordinates with respect
to some conveniently defined stationary frame, $\mu$ refer to co-ordinates with respect to the Fermi frame, and $\hat{\mu}$ denote
co-ordinates for a local Lorentz frame at the worldline.
Spatial co-ordinates are labelled from $1$ to $3$ by Latin indices.

To begin, consider a worldline ${\cal C}$ in a general space-time background parametrized by proper time $\tau$.
At some event $P_0$ on ${\cal C}$, the Fermi frame \cite{Manasse,Synge,Mashhoon1,Mashhoon2} is determined by
constructing a local orthonormal vierbein set $\lt\{ \lm^{\bar{\alpha}}{}_\mu \rt\} \,$ and its inverse set $\lt\{ \lm^\mu{}_{\bar{\alpha}} \rt\}$,
such that $\lm^{\bar{\mu}}{}_0 = \d x^{\bar{\mu}}/\d \tau$ and $\lm^{\bar{\mu}}{}_a$ define the local spatial axes.
If the unit spatial tangent vector from $P_0$ to a neighbouring event $P$ is described by $\xi^{\bar{\mu}} = \lt(\d x^{\bar{\mu}}/\d \sg\rt)_0$,
where $\sg$ is the proper length of a unique spacelike geodesic orthogonal to ${\cal C}$, then the Fermi normal co-ordinates at $P$ are described by
$T = \tau$ and $X^i = \sg \, \xi^{\bar{\mu}} \, \lm^i{}_{\bar{\mu}}$.
It follows that the space-time metric is described by
\be
\d s^2 & = & {}^F{}g_{\mu \nu}(X) \, \d X^\mu \, \d X^\nu \, ,
\label{F-metric}
\ee
where
%
\be
{}^F{}g_{00}(X) & = & 1 + {}^F{}R_{0i0j}(T) \, X^i \, X^j + \cdots \, ,
\label{F-g00}
\nl
{}^F{}g_{0j}(X) & = & {2 \over 3} \, {}^F{}R_{0ijk}(T) \, X^i \, X^k + \cdots \, ,
\label{F-g0j}
\nl
{}^F{}g_{ij}(X) & = & \eta_{ij} + {1 \over 3} \, {}^F{}R_{ikjl}(T) \, X^k \, X^l + \cdots \, ,
\label{F-gij}
\ee
%
and
\be
{}^F{}R_{\al \bt \gm \dl}(T) & = & R_{\bar{\mu} \bar{\nu} \bar{\rho} \bar{\sg}} \,
\lm^{\bar{\mu}}{}_\al \, \lm^{\bar{\nu}}{}_\bt \, \lm^{\bar{\rho}}{}_{\gm} \, \lm^{\bar{\sg}}{}_\dl \, .
\label{F-Riemann}
\ee
It is possible to find an orthonormal vierbein set $\lt\{ \bar{e}^\mu{}_{\hat{\alpha}}(X) \rt\}$ and its inverse
$\lt\{ \bar{e}^{\hat{\alpha}}{}_\mu (X)\rt\} \,$ to define a local Lorentz frame \cite{Poisson} satisfying
${}^F{}g_{\mu \nu}(X) = \eta_{\hat{\alpha}\hat{\beta}} \, \bar{e}^{\hat{\alpha}}{}_\mu (X) \, \bar{e}^{\hat{\beta}}{}_\nu (X)$,
where
%
\be
\bar{e}^{\hat{0}}{}_0(X) & = & 1 + {1 \over 2} \, {}^F{}R_{0i0j}(T) \, X^i \, X^j \, ,
\label{inv-tetrad-00}
\nl
\bar{e}^{\hat{0}}{}_j(X) & = & {1 \over 6} \, {}^F{}R_{0ijk}(T) \, X^i \, X^k \, ,
\label{inv-tetrad-0j}
\nl
\bar{e}^{\hat{\imath}}{}_0(X) & = & -{1 \over 2} \, {}^F{}R^i{}_{j0k}(T) \, X^j \, X^k \, ,
\label{inv-tetrad-i0}
\nl
\bar{e}^{\hat{\imath}}{}_j(X) & = & \dl^i{}_j - {1 \over 6} \, {}^F{}R^i{}_{kjl}(T) \, X^k \, X^l \, ,
\label{inv-tetrad-ij}
\ee
%
and
%
\be
\bar{e}^0{}_{\hat{0}}(X) & = & 1 - {1 \over 2} \, {}^F{}R_{0i0j}(T) \, X^i \, X^j \, ,
\label{tetrad-00}
\nl
\bar{e}^i{}_{\hat{0}}(X) & = & {1 \over 2} \, {}^F{}R^i{}_{j0k}(T) \, X^j \, X^k \, ,
\label{tetrad-i0}
\nl
\bar{e}^0{}_{\hat{\jmath}}(X) & = & -{1 \over 6} \, {}^F{}R_{0ijk}(T) \, X^i \, X^k \, ,
\label{tetrad-0j}
\nl
\bar{e}^i{}_{\hat{\jmath}}(X) & = & \dl^i{}_j + {1 \over 6} \, {}^F{}R^i{}_{kjl}(T) \, X^k \, X^l \, .
\label{tetrad-ij}
\ee
%

\subsubsection{Conversion to Curvilinear Co-ordinates}
Because the spatial components of the Fermi normal co-ordinates are locally Cartesian given (\ref{F-metric}),
the underlying issue remains as before \cite{Singh-Mobed}, namely whether the propagation of a quantum particle
in a curved background is properly represented by $X^\mu$ or if the spatial components should be generalized to accommodate
the symmetries of the particle's motion in space.
To address this issue, it is necessary to allow $X^i = X^i\lt(u^1, u^2, u^3\rt)$, where $u^k$ are general
curvilinear co-ordinates \cite{Singh-Mobed2}.
This leads to a new expression of Fermi normal co-ordinates in the form $U^\mu$, where $U^0 = T$ and $U^i = u^i$.
It clearly follows that a new set of orthonormal vierbeins for these new co-ordinates are expressed as
%
\be
e^\bt{}_{\hat{\al}}(U) & = & {\partial U^\bt \over \partial X^\al} \, \bar{e}^\al{}_{\hat{\al}}(X) \, , \qquad
\label{tetrad-coord-trans}
\nl
e^{\hat{\al}}{}_\bt(U) & = & {\partial X^\al \over \partial U^\bt} \, \bar{e}^{\hat{\al}}{}_\al(X) \, .
\label{inv-tetrad-coord-trans}
\ee
%
Then Minkowski space-time in curvilinear co-ordinate form is described by \cite{Hassani}
\be
\d s^2 & = & \d T^2 + \eta_{\hat{\imath}\hat{\jmath}} \lt(\lm^{(\hat{\imath})}(u) \, \d u^{\hat{\imath}}\rt)
\lt(\lm^{(\hat{\jmath})}(u) \, \d u^{\hat{\jmath}}\rt) \, ,
\label{Minkowski-curvilinear}
\ee
with $\lm^{(\hat{\imath})}(u)$ as dimensional scale functions.

\subsection{Covariant Dirac Equation}

Given (\ref{F-g00})--(\ref{F-gij}), the covariant Dirac equation for a spin-1/2 particle with mass $m$ can be written as
\be
\lt[i \gm^\mu(X) \lt[\partial_\mu + i \, \Gamma_\mu(X) \rt] - m/\hb\rt]\psi(X) & = & 0 \, ,
\label{Dirac-eq}
\ee
where $\partial_\mu = \partial/\partial X^\mu$,
$\lt\{ \gm^\mu(X) \rt\}$ is the set of gamma matrices satisfying
$\lt\{ \gm^\mu(X), \gm^\nu(X) \rt\} = 2 \, g_F^{\mu \nu}(X)$, and $\Gamma_\mu(X)$ is the spin connection.
With (\ref{inv-tetrad-00})--(\ref{tetrad-ij}) available, the spin connection can be expressed as
\be
\Gm_\mu(X) & = & {i \over 4} \, \gm^\al(X) \lt(\nabla_\mu \gm_\al(X) \rt)
\ = \ -{1 \over 4} \, \sg^{\hat{\al} \hat{\bt}} \, \eta_{\hat{\bt} \hat{\gm}} \, \bar{e}^\al{}_{\hat{\al}} \lt(\nabla_\mu \, \bar{e}^{\hat{\gm}}{}_\al \rt),
\label{spin-connection}
\ee
where $\nabla_\mu$ denotes covariant differentiation, $\lt\{ \gm^{\hat{\al}} \rt\}$ is the set of flat space-time gamma matrices satisfying
$\lt\{ \gm^{\hat{\al}}, \gm^{\hat{\bt}} \rt\} = 2 \, \eta^{\hat{\al} \hat{\bt}}$,
following the convention of Itzykson and Zuber \cite{Itzykson}, and $\sg^{\hat{\al} \hat{\bt}} = {i \over 2} \lt[\gm^{\hat{\al}}, \gm^{\hat{\bt}}\rt]$.
Retaining only contributions to first-order in the Riemann tensor, it therefore follows from (\ref{spin-connection}) that
\be
\Gm_0(X) & = & i \, \gm^{\hat{0}} \, \gm^{\hat{\imath}} \lt[{1 \over 2} \, {}^F{}R_{i00k}(T) + {1 \over 3} \, {}^F{}R_{ij0k,0}(T) \, X^j \rt] X^k \, ,
\label{Gm-0}
\nl
\Gm_l(X) & = & i \, \gm^{\hat{0}} \, \gm^{\hat{\imath}} \lt[{1 \over 3} \, \lt({}^F{}R_{il0k}(T) + {}^F{}R_{i[k0]l}(T)\rt)
- {1 \over 12} \, {}^F{}R_{ijkl,0}(T) \, X^j \rt] X^k \, , \qquad
\label{Gm-l}
\ee
where ${}^F{}R_{i[k0]l}(T) = {1 \over 2} \lt[{}^F{}R_{ik0l}(T) - {}^F{}R_{i0kl}(T)\rt]$.
The spin connection in curvilinear co-ordinates is then
\be
\Gm_\mu(U) & = & {i \over 4} \, \gm^\al(U) \lt[\nabla_\mu \gm_\al(U) \rt]
\ = \ -{1 \over 4} \, \sg^{\hat{\al} \hat{\bt}} \, e^{\hat{\mu}}{}_\mu \, \Om_{\hat{\al} \hat{\bt} \hat{\mu}}
\nn
& = & {\partial X^\al \over \partial U^\mu} \, \Gm_\al(X) \, ,
\label{spin-connection-curvilinear}
\ee
where
\be
\Om_{\hat{\al} \hat{\bt} \hat{\mu}}(U) & = &
- \eta_{\hat{\gm} [\hat{\al}} \, e^\nu{}_{\hat{\bt}]} \lt(\nabla_{\hat{\mu}} \, e^{\hat{\gm}}{}_\nu \rt) \,
\nn
& = & - \eta_{\hat{\gm} [\hat{\al}} \, \bar{e}^\nu{}_{\hat{\bt}]} \lt(\nabla_{\hat{\mu}} \, \bar{e}^{\hat{\gm}}{}_\nu \rt)
\ = \ \Om_{\hat{\al} \hat{\bt} \hat{\mu}}(X) \,
\label{Omega-3-indices}
\ee
is the Ricci rotation coefficient \cite{Chandra}.

By projecting onto the local Lorentz frame, the covariant Dirac equation (\ref{Dirac-eq}) expressed in curvilinear co-ordinates
becomes
\be
\lt[i \gm^{\hat{\mu}} \lt(\hat{\nabla}_{\hat{\mu}} + i \, \Gm_{\hat{\mu}}(U) \rt) - m/\hb\rt]\psi(U) & = & 0 \, ,
\label{Dirac-eq-curvilinear}
\ee
where $\hat{\nabla}_{\hat{\mu}} = \nabvec_{\hat{\mu}} + i \, \hat{\Gm}_{\hat{\mu}}(U)$ is the covariant derivative operator
with
\be
\nabvec_{\hat{0}} & \equiv & {\partial \over \partial T} \, , \qquad
\nabvec_{\hat{\jmath}} \ \equiv \ {1 \over \lm^{(\hat{\jmath})}(u)} \, {\partial \over \partial u^{\hat{\jmath}}} \, .
\label{nabvec}
\ee
It is fairly lengthy but straightforward to show that (\ref{Dirac-eq-curvilinear}) takes the form
\be
\lt[\gm^{\hat{\mu}} \lt(\Pvec_{\hat{\mu}} - \hb \, \Gmvec_{\hat{\mu}}(U) \rt) - m\rt]\psi(U) & = & 0 \, ,
\label{Dirac-eq-curvilinear-1}
\ee
where
\be
\Pvec_{\hat{\mu}} & = & \pvec_{\hat{\mu}} + \Omvec_{\hat{\mu}}
\label{Pvec}
\ee
is the momentum operator in curvilinear co-ordinates, in terms of
%
\be
\pvec_{\hat{\mu}} & = & i \hbar \, \nabvec_{\hat{\mu}} \, ,
\label{pvec}
\nl
\Omvec_{\hat{\mu}} & = & i \hbar \lt[\nabvec_{\hat{\mu}} \ln \lt(\lm^{(\hat{1})}(u) \, \lm^{(\hat{2})}(u) \, \lm^{(\hat{3})}(u)\rt)^{1/2} \rt] \, ,
\label{Ovec}
\nl
i \, \Gmvec_{\hat{\mu}} & = & \bar{\Gmvec}^{(\rm S)}_{\hat{\mu}}
+ \gm^{\hat{l}} \, \gm^{\hat{m}} \, \bar{\Gmvec}^{(\rm T)}_{\ze[\hat{l}\hat{m}]} \, \dl^{\ze}{}_{\hat{\mu}} \, ,
\label{Spin-Connection}
\ee
%
where
%
\be
\bar{\Gmvec}^{(\rm S)}_{\ze} & = & {1 \over 12} {}^F{}R^m{}_{jmk,0}(T) \, X^j \, X^k \, ,
\label{Spin-Connection-0-S}
\nl
\bar{\Gmvec}^{(\rm S)}_{\hat{\jmath}} & = & - \lt[{1 \over 2} {}^F{}R_{j00m}(T) + {1 \over 3} {}^F{}R_{jl0m,0}(T) \, X^l \rt] X^m \, ,
\qquad \hspace{1mm}
\label{Spin-Connection-j-S}
\nl
\bar{\Gmvec}^{(\rm T)}_{\ze[\hat{l}\hat{m}]} & = & 
{1 \over 2} \, {}^F{}R_{lm0k}(T) \, X^k \, . \qquad
\label{Spin-Connection-0-T}
\ee
%
It is easy to verify for spherical co-ordinates $u^k = \lt(r, \th, \ph\rt)$ that \cite{Singh-Mobed}
\be
\Pvec^{\hat{r}} & = & - \ihb \lt(\pt{r} + {1 \over r}\rt) \, , \quad
\Pvec^{\hat{\th}} \ = \ - {\ihb \over r} \lt(\pt{\th} + {1 \over 2} \, \cot \th\rt) \, ,
\nn
\Pvec^{\hat{\ph}} & = & - {\ihb \over r \, \sin \th} \, \pt{\ph} \, ,
\label{momentum-spherical}
\ee
where $\lm^{(\hat{1})}(u) = 1 \,$, $\lm^{(\hat{2})}(u) = r \,$, and $\lm^{(\hat{3})}(u) = r \, \sin \th \,$.
By letting $\Dvec_{\hat{\mu}} = \Pvec_{\hat{\mu}} - \hb \, \Gmvec_{\hat{\mu}}$ and recalling the identity \cite{Aitchison}
\be
\gm^{\hat{\mu}} \, \gm^{\hat{\nu}} \, \gm^{\hat{\rho}} & = & \eta^{\hat{\nu} \hat{\rho}} \, \gm^{\hat{\mu}}
- 2 \, \gm^{[\hat{\nu}} \eta^{\hat{\rho}]\hat{\mu}} - i \, \gm^5 \, \gm^{\hat{\sg}} \, \varepsilon^{\hat{\mu} \hat{\nu} \hat{\rho}}{}_{\hat{\sg}} \, ,
\label{gm-identity}
\ee
where $\varepsilon^{\hat{\mu} \hat{\nu} \hat{\rho} \hat{\sg}}$ is the Levi-Civita symbol \cite{Itzykson} with
$\varepsilon^{\hat{0} \hat{1} \hat{2} \hat{3}} = 1$, it is shown that
\be
\Dvec_{\hat{\mu}}^{(1/2)} & = & \Pvec_{\hat{\mu}} -
\hb \lt(\gm^5 \, \bar{\Gmvec}^{(\rm C)}_{\hat{\mu}} - i \, \bar{\Gmvec}^{(\rm S)}_{\hat{\mu}}\rt) \, ,
\label{D}
\nl
\bar{\Gmvec}^{(\rm C)}_{\hat{\mu}} & = & \varepsilon^{\hat{0} \hat{l} \hat{m}}{}_{\hat{\mu}} \, \bar{\Gmvec}^{(\rm T)}_{\ze[\hat{l}\hat{m}]} \, .
\label{Spin-Connection-C}
\ee
The ``C'' in (\ref{Spin-Connection-C}) denotes the chiral-dependent part of the spin connection, while the ``S'' in
(\ref{Spin-Connection-0-S})--(\ref{Spin-Connection-j-S}) refers to the symmetric part under chiral symmetry.
Finally, for the purpose of generalization, assume that the elementary particle is
sensitive to electromagnetic fields, and that
$\Pvec_{\hat{\mu}} \rightarrow \bar{\Pvec}_{\hat{\mu}} = \Pvec_{\hat{\mu}} - e \, \Avec_{\hat{\mu}}$,
where $\Avec_{\hat{\mu}}$ is the electromagnetic vector potential and $e$ is the particle's unit charge.

\subsection{Casimir Scalars for Linear Momentum and Spin Angular Momentum}

For spin-1/2 particles, it is straightforward to show that the Casimir scalar for linear momentum is
\be
\Dvec^{\hat{\al}}_{(1/2)} \, \Dvec_{\hat{\al}}^{(1/2)} & = & m_0^2
- 2 \, \hbar \lt(\gm^5 \, \bar{\Gmvec}^{(\rm C)}_{\hat{\al}} - i \, \bar{\Gmvec}^{(\rm S)}_{\hat{\al}}\rt) \bar{\Pvec}^{\hat{\al}}
- \hbar^2 \, \nabvec^{\hat{\al}} \lt(\bar{\Gmvec}^{(\rm S)}_{\hat{\al}} + i \, \gm^5 \, \bar{\Gmvec}^{(\rm C)}_{\hat{\al}}\rt) ,
\label{D^2-spin-1/2}
\ee
where $m_0^2 \equiv \bar{\Pvec}^{\hat{\al}} \, \bar{\Pvec}_{\hat{\al}}$ and
$m_0^2 = m^2$ in the absence of external fields.
It becomes self-evident that (\ref{D^2-spin-1/2}) is a Lorentz invariant under these conditions.

To determine the Casimir scalar for spin angular momentum, recall that the Pauli-Lubanski vector is
formally defined in a local Lorentz frame as
\be
\Wvec^{\hat{\mu}} & = & -{1 \over 2} \, \varepsilon^{\hat{\mu}}{}_{\hat{\al}\hat{\bt}\hat{\gm}} \,
\Sgvec^{\hat{\al}\hat{\bt}} \, \Dvec^{\hat{\gm}} \, .
\label{Wvec-defn}
\ee
For spin-1/2 particles, $\Sgvec_{(1/2)}^{\hat{\al}\hat{\bt}} = \sg^{\hat{\al}\hat{\bt}}/2$ is identified
in terms of the local Lorentz transformation for a spinor field $\Psi_{(1/2)}$, such that \cite{Birrell}
\be
\Psi'_{(1/2)} & = & \lt(1 - {i \over 2} \, \om^{\hat{\al} \hat{\bt}} \, \Sgvec_{\hat{\al} \hat{\bt}}^{(1/2)}\rt) \Psi_{(1/2)} \, ,
\label{spin-1/2}
\ee
with $\om^{\hat{\al} \hat{\bt}}$ denoting infinitesimal rotations in space-time.
By using (\ref{gm-identity}) and the identity
$\varepsilon_{\hat{\mu} \hat{\nu} \hat{\rho} \hat{\sg}} \, \sg^{\hat{\rho} \hat{\sg}}
= -2 \, i \, \gm^5 \, \sg_{\hat{\mu} \hat{\nu}}$ \cite{Itzykson}, it follows that
\be
\Wvec^{\hat{\al}}_{(1/2)} \, \Wvec_{\hat{\al}}^{(1/2)} & = &
-{3 \over 4} \, \Dvec^{\hat{\al}}_{(1/2)} \, \Dvec_{\hat{\al}}^{(1/2)} 
+ {i \over 4} \, \sg^{\hat{\al}\hat{\bt}} \lt[\Dvec_{\hat{\al}}^{(1/2)} , \Dvec_{\hat{\bt}}^{(1/2)}\rt] \, , \hspace{1mm}
\label{W^2-spin-1/2-formal}
\ee
where the first term of (\ref{W^2-spin-1/2-formal}) is identified with $-{1 \over 2}\lt({1 \over 2} + 1 \rt) m^2$ in flat space-time
without external fields.
As for the second term of (\ref{W^2-spin-1/2-formal}), this is determined according to \cite{Pagels}
%
\be
i \lt[\Dvec_{\hat{\al}}^{(1/2)} , \Dvec_{\hat{\bt}}^{(1/2)} \rt] & = & {\hbar^2 \over 4} \, \sg^{\hat{\mu}\hat{\nu}}
\lt( {}^F{}R_{\hat{\mu}\hat{\nu}\hat{\al}\hat{\bt}} \rt) - e \, \hbar \lt( {}^F{}F_{\hat{\al} \hat{\bt}} \rt)
- \hbar \, C^{\hat{\mu}}{}_{\hat{\al}\hat{\bt}} \lt(e \, \Avec_{\hat{\mu}} + \hbar \, \Gmvec_{\hat{\mu}} \rt)
\nn
&  &{} + i \lt[\Pvec_{\hat{\al}} , \Pvec_{\hat{\bt}}\rt] \, ,
\label{D-commutator}
\nl
i \lt[\Pvec_{\hat{\al}} , \Pvec_{\hat{\bt}}\rt] & = & \hbar \, C^{\hat{\mu}}{}_{\hat{\al}\hat{\bt}} \, \Pvec_{\hat{\mu}}
+ 2 \, \hbar \, \bar{e}^\sg{}_{[\hat{\al}} \, \bar{e}^\gm{}_{\hat{\bt}]} \lt(\nabvec_{\hat{\sg}} \, \ln \lm^{(\gm)}\rt) \Pvec_{\hat{\gm}}
 \, ,
\qquad \hspace{2mm}
\label{P-commutator}
\nl
C^{\hat{\mu}}{}_{\hat{\al}\hat{\bt}} & = & 
2 \, \bar{e}^{\hat{\mu}}{}_\sg \lt( \nabla_\lm \, \bar{e}^\sg{}_{[{\hat{\al}}} \rt) \bar{e}^\lm{}_{{\hat{\bt}}]} \, ,
\label{C-defn}
\ee
%
where $C^{\hat{\mu}}{}_{\hat{\al}\hat{\bt}}$ is the exclusively curvature-dependent object of anholonomicity
for the local Lorentz frame.
The second term in (\ref{P-commutator}) contains the non-inertial dipole operator, given by
\be
\Rvec^{\hat{k}} & = & \lt. {i \over 2\hbar} \, \varepsilon^{\hat{0} \hat{\imath} \hat{\jmath} \hat{k}} \,
[\Pvec_{\hat{\imath}}, \Pvec_{\hat{\jmath}}] \rt|_{{}^F{}R_{\hat{\mu}\hat{\nu}\hat{\al}\hat{\bt}} \rightarrow 0}
\ = \ \varepsilon^{\hat{0} \hat{\imath} \hat{\jmath} \hat{k}} \,
\lt(\nabvec_{\hat{\imath}} \, \ln \lm^{(j)}\rt) \Pvec_{\hat{\jmath}} \, .
\label{Rvec}
\ee
The existence of the second term in (\ref{P-commutator}) is a subtle issue that requires justification,
since the standard expression of $i \lt[\Pvec_{\hat{\al}} , \Pvec_{\hat{\bt}}\rt] =
\hbar \, C^{\hat{\mu}}{}_{\hat{\al}\hat{\bt}} \, \Pvec_{\hat{\mu}}$ is what usually appears in the literature.
A derivation of (\ref{P-commutator}) can be found in \ref{Appendix-A} of this paper.

Substituting (\ref{D-commutator})--(\ref{Rvec}) into (\ref{W^2-spin-1/2-formal}) leads to
\be
\Wvec^{\hat{\al}}_{(1/2)} \, \Wvec_{\hat{\al}}^{(1/2)} & = &
-{1 \over 2} \lt({1 \over 2} + 1\rt) \lt[m_0^2 + {\hbar^2 \over 6} \lt({}^F{}R^{\hat{\al}}{}_{\hat{\al}}\rt) \rt]
+ {\hbar \over 2} \lt(\sgvec \cdot \Rvec\rt)
\nn
&  &{} - {\hbar \over 4} \, \sg^{\hat{\al} \hat{\bt}}
\lt[Q_{\hat{\al}\hat{\bt}} - e \lt({}^F{}F_{\hat{\al} \hat{\bt}} + C^{\hat{\mu}}{}_{\hat{\al}\hat{\bt}} \, \Avec_{\hat{\mu}}\rt) \rt]
\nn
&  &{} + {3 \over 2} \, \hbar \lt(\gm^5 \, \bar{\Gmvec}^{(\rm C)}_{\hat{\al}} - i \, \bar{\Gmvec}^{(\rm S)}_{\hat{\al}}\rt) \bar{\Pvec}^{\hat{\al}}
+ {3 \over 4} \, \hbar^2 \, \nabvec^{\hat{\al}}
\lt(\bar{\Gmvec}^{(\rm S)}_{\hat{\al}} + i \, \gm^5 \, \bar{\Gmvec}^{(\rm C)}_{\hat{\al}}\rt) \,
\label{W^2-spin-1/2}
\ee
as the expression for the spin-1/2 particle Casimir scalar for spin, where
\be
Q_{\hat{\al}\hat{\bt}} & = &
{i \over \hbar} \lt[\Pvec_{\hat{\al}} , \Pvec_{\hat{\bt}}\rt]
- \dl^{\hat{\imath}}{}_{\hat{\al}} \, \dl^{\hat{\jmath}}{}_{\hat{\bt}} \,
\varepsilon_{\ze {\hat{\imath}} {\hat{\jmath}} {\hat{k}}} \, \Rvec^{\hat{k}} \,
\nn
& = & 2 \lt(\bar{e}^\sg{}_{[\hat{\al}} \, \bar{e}^\gm{}_{\hat{\bt}]} - \dl^\sg{}_{[\hat{\al}} \, \dl^\gm{}_{\hat{\bt}]}\rt)
\lt(\nabvec_{\hat{\sg}} \, \ln \lm^{(\gm)}\rt) \Pvec_{\hat{\gm}}
+ C^{\hat{\mu}}{}_{\hat{\al}\hat{\bt}} \, \Pvec_{\hat{\mu}} \,
\label{Q-defn}
\ee
is first-order in the Riemann tensor and $Q_{\hat{\al}\hat{\bt}} \rightarrow 0$ as
${}^F{}R_{\hat{\mu}\hat{\nu}\hat{\al}\hat{\bt}} \rightarrow 0$.
It is clear to see that (\ref{W^2-spin-1/2}) reduces to the frame-dependent expression obtained
in flat space-time when all external fields vanish \cite{Singh-Mobed}.
As well, it is straightforward to show from the contraction of $\Wvec^{(1/2)}$ defined by (\ref{Wvec-defn}) with $\Dvec^{(1/2)}$
that the orthogonality relationship between the two vectors is no longer satisfied, since
\be
\Wvec_{(1/2)}^{\hat{\mu}} \, \Dvec^{(1/2)}_{\hat{\mu}} & = & - \Dvec_{(1/2)}^{\hat{\mu}} \, \Wvec^{(1/2)}_{\hat{\mu}}
\nn
& = & - \lt\{ {\hbar \over 2} \lt(\alvec \cdot \Rvec\rt)
+ {\hbar \over 4} \gm^5 \, \sg^{\hat{\al} \hat{\bt}}
\lt[Q_{\hat{\al}\hat{\bt}} - e \lt({}^F{}F_{\hat{\al} \hat{\bt}} + C^{\hat{\mu}}{}_{\hat{\al}\hat{\bt}} \, \Avec_{\hat{\mu}}\rt) \rt. \rt.
\nn
&  &{} - \lt. \lt. \hbar \, C^{\hat{\mu}}{}_{\hat{\al}\hat{\bt}}
\lt(\gm^5 \, \bar{\Gmvec}^{(\rm C)}_{\hat{\mu}} - i \, \bar{\Gmvec}^{(\rm S)}_{\hat{\mu}}\rt) \rt]
+ {\hbar^2 \over 8} \, \gm^5 \lt({}^F R^{\hat{\al}}{}_{\hat{\al}} \rt) \rt\} \, ,
\label{W.D-spin-1/2}
\ee
which also reduces to its flat space-time counterpart \cite{Singh-Mobed}.
Since this holds true even for {\em massless} $(m_0 = 0)$ spin-1/2 particles in the absence of external fields,
the standard identification of helicity state projections parallel and antiparallel to the particle's
direction of motion \cite{Ryder} no longer applies when considering non-inertial motion with rotation.
It is also interesting to note that, for the chiral representation \cite{Itzykson} with $\alvec = \gm^5 \, \sgvec$,
almost all the Hermitian terms in (\ref{W.D-spin-1/2}) exist as pseudo-scalars.

\section{Casimir Invariance Breakdown due to Non-Inertial Motion}
\label{Sec:Casimir-violation}

The scalars derived from the momentum operator $\Dvec_{\hat{\mu}}$ and the spin operator $\Wvec_{\hat{\mu}}$ are purposefully
labelled ``Casimir scalars'' instead of ``Casimir invariants'' to underscore the fact that at least
$\Wvec^{\hat{\mu}} \, \Wvec_{\hat{\mu}}$ is no longer a Lorentz invariant quantity when considering free-particle non-inertial motion
while in a flat background.
However, by definition a Casimir operator is an element in a Lie algebra that commutes with
{\em all} of the generators that define the algebra \cite{Gursey}.
If the spatial canonical momentum generators are described by curvilinear co-ordinates, it becomes evident
that even this designation of ``Casimir scalars'' for $\Dvec^{\hat{\mu}} \, \Dvec_{\hat{\mu}}$ and
$\Wvec^{\hat{\mu}} \, \Wvec_{\hat{\mu}}$ is somewhat misleading, since in the absence of external fields,
%
\be
i \lt[\lt(\Pvec^{\hat{\mu}} \, \Pvec_{\hat{\mu}}\rt) , \, \Pvec_{\hat{\nu}} \rt]
& = & 2 \hbar \, \varepsilon_{\hat{0} \hat{\imath} \hat{\jmath} \hat{k}}
\lt[\Rvec^{\hat{\imath}} \, \Pvec^{\hat{\jmath}} + {i \over 2} \lt(\nabvec^{\hat{\jmath}} \Rvec^{\hat{\imath}}\rt) \rt]
\dl^{\hat{k}}{}_{\hat{\nu}} \, ,
\label{Casimir-momentum-violation}
\nl
\nn
i \lt[\lt(\Wvec^{\hat{\mu}} \, \Wvec_{\hat{\mu}}\rt) , \, \Pvec_{\hat{\nu}} \rt]
& = & 2 i \lt[\Wvec_{\hat{\mu}} \, , \, \Pvec_{\hat{\nu}} \rt]
\Wvec^{\hat{\mu}} + {\hbar \over 2} \, \varepsilon^{\hat{\mu}}{}_{\hat{\al} \hat{\bt} \hat{\gm}} \,
\Sgvec^{\hat{\al} \hat{\bt}} \lt(\nabvec^{\hat{\gm}} \lt[\Wvec_{\hat{\mu}} \, , \, \Pvec_{\hat{\nu}} \rt]\rt)
\, ,
\label{Casimir-spin-violation}
\ee
%
where
\be
i \lt[\Wvec_{\hat{\mu}} \, , \, \Pvec_{\hat{\nu}} \rt] & = & -2 \hbar \lt(\eta_{{\hat{\mu}} [\hat{0}} \, \Sgvec_{\hat{\jmath}] {\hat{\nu}}}
+ {1 \over 2} \, \eta_{{\hat{\mu}} {\hat{\nu}}} \, \Sgvec_{\hat{0} \hat{\jmath}}\rt) \Rvec^{\hat{\jmath}} \, .
\label{[W,P]}
\ee
It is obvious that $\Pvec^{\hat{\mu}} \, \Pvec_{\hat{\mu}}$ and $\Wvec^{\hat{\mu}} \, \Wvec_{\hat{\mu}}$
become Casimir invariants in general {\em only} when $\Rvec \rightarrow \vec{0}$, corresponding to rectilinear motion.
Nonetheless, for the wider purposes of this paper and for the sake of convenience, these operators are identified as
``Casimir scalars'' for the remainder of this paper, while clearly acknowledging the conceptual disagreement
between this designation and the formal definition to which it refers \cite{Gursey}.

\section{Casimir Scalars for Particles of Higher-Order Spin}
\label{Sec.Casimir.higher-spin}

The basis for exploring the Casimir properties of integer and half-integer spin elementary particles is
given by the local Lorentz transformation of a general multi-indexed field $T^{\hat{\nu}_1 \cdots \hat{\nu}_N}_{(s)}$
of spin-$s$, such that \cite{Birrell}
\be
T'^{\hat{\mu}_1 \cdots \hat{\mu}_N}_{(s)} & = &
\lt(1 - {i \over 2} \, \om^{\hat{\al} \hat{\bt}} \, \Sgvec_{\hat{\al} \hat{\bt}}^{(s)}
\rt)^{\hat{\mu}_1 \cdots \hat{\mu}_N}{}_{\hat{\nu}_1 \cdots \hat{\nu}_N} \, T^{\hat{\nu}_1 \cdots \hat{\nu}_N}_{(s)} \, ,
\label{spin-general}
\ee
where $\lt[\Sgvec_{\hat{\al} \hat{\bt}}^{(s)}\rt]^{\hat{\mu}_1 \cdots \hat{\mu}_N}{}_{\hat{\nu}_1 \cdots \hat{\nu}_N}$
is the respective covariant spin operator.
When $N = 0$, it follows that $T^{\hat{\nu}_1 \cdots \hat{\nu}_N}_{(s)}$ is the spin-1/2 field $\Psi_{(1/2)}$,
while $N > 0$ corresponds to higher-order spin fields.
Then the covariant momentum operator $\Dvec_{\hat{\mu}}^{(s)} = \bar{\Pvec}_{\hat{\mu}} - \hbar \, \Gmvec_{\hat{\mu}}^{(s)}$
is described for spin-$s$ by \cite{Birrell}
\be
\lt[\Dvec_{\hat{\mu}}^{(s)}\rt]^{\hat{\mu}_1 \cdots \hat{\mu}_N}{}_{\hat{\nu}_1 \cdots \hat{\nu}_N}
& = & \lt(\bar{\Pvec}_{\hat{\mu}} + {\hbar \over 2} \, \Sgvec^{\hat{\al} \hat{\bt}}_{(s)} \,
\Om_{\hat{\al} \hat{\bt} \hat{\mu}}\rt)^{\hat{\mu}_1 \cdots \hat{\mu}_N}{}_{\hat{\nu}_1 \cdots \hat{\nu}_N} \, ,
\label{D-general}
\ee
while the corresponding expression for the Pauli-Lubanski vector is
\be
\lt[\Wvec^{\hat{\mu}}_{(s)}\rt]^{\hat{\mu}_1 \cdots \hat{\mu}_N}{}_{\hat{\nu}_1 \cdots \hat{\nu}_N} & = &
\lt(-{1 \over 2} \, \varepsilon^{\hat{\mu}}{}_{\hat{\al}\hat{\bt}\hat{\gm}} \,
\Sgvec^{\hat{\al}\hat{\bt}}_{(s)} \, \Dvec^{\hat{\gm}}_{(s)}\rt)^{\hat{\mu}_1 \cdots \hat{\mu}_N}{}_{\hat{\nu}_1 \cdots \hat{\nu}_N} \, .
\label{W-general}
\ee
With (\ref{D-general}) and (\ref{W-general}), it becomes a straightforward exercise to determine the Casimir scalars
$\Dvec^{(s)} \cdot \Dvec^{(s)}$ and $\Wvec^{(s)} \cdot \Wvec^{(s)}$ for higher-order spin.
The computations for spin-1, spin-3/2, and spin-2 are presented below, where the spin-3/2 case is described
by the Rarita-Schwinger vector-spinor field \cite{Rarita}.
All the fields considered are assumed to be massive $(m_0 \neq 0)$, where the special case of massless fields
$(m_0 = 0)$ is treated separately.
With the exception of \ref{Appendix-A}, the orthonormal frame indices for the remainder of this paper will be
unhatted for the sake of notational clarity.

\subsection{Spin-1 Fields}

The local Lorentz transformation for a spin-1 vector field $V^\mu_{(1)}$ is given by
\be
V'^\mu_{(1)} & = & \lt(\dl^\mu{}_\nu
- {i \over 2} \, \om^{\al \bt} \lt[\Sgvec_{\al \bt}^{(1)}\rt]^\mu{}_\nu \rt) V^\nu_{(1)} \, ,
\label{spin-1}
\ee
where the covariant spin operator $\Sgvec_{\al \bt}^{(1)}$ is
\be
\lt[\Sgvec_{\al \bt}^{(1)}\rt]^\mu{}_\nu & = & 2i \, \dl^\mu{}_{[\al} \eta_{\bt] \nu} \, .
\label{om-spin-1}
\ee
Then the combination of (\ref{D-general}) and (\ref{om-spin-1}) leads to
\be
\lt[\Dvec_\mu^{(1)}\rt]^\lm{}_\rho & = & \dl^\lm{}_\rho \, \bar{\Pvec}_\mu
+ {\hbar \over 2} \lt[\Sgvec^{\al \bt}_{(1)}\rt]^\lm{}_\rho \, \Om_{\al \bt \mu}
\ = \ \dl^\lm{}_\rho \, \bar{\Pvec}_\mu + i \hbar \, \Om^\lm{}_{\rho \mu} \,
\label{D-spin-1}
\ee
for the spin-1 covariant momentum operator, while (\ref{om-spin-1}) and (\ref{D-spin-1}) substituted into
(\ref{W-general}) results in
\be
\lt[\Wvec^\mu_{(1)} \rt]^\lm{}_\rho & = & - {1 \over 2} \, \varepsilon^{\mu \al \bt \gm}
\, \lt[\Sgvec_{\al \bt}^{(1)}\rt]^\lm{}_\sg \, \lt[\Dvec_\gm^{(1)}\rt]^\sg{}_\rho
\nn
& = & -i \, \varepsilon^{\mu \lm}{}_{\rho \gm} \, \bar{\Pvec}^\gm
- \hbar \, \varepsilon^{\mu \lm \al \bt} \, \Om_{\rho \al \bt} \,
\label{W-spin-1}
\ee
for the spin-1 Pauli-Lubanski vector.

A straightforward calculation of $\Dvec^{(1)} \cdot \Dvec^{(1)}$ from (\ref{D-spin-1}) leads to
\be
\lt[\Dvec^\al_{(1)} \, \Dvec_\al^{(1)} \rt]^\mu{}_\nu & = &
m_0^2 \, \dl^\mu{}_\nu + 2i \hbar \, \Om^\mu{}_{\nu \al} \, \bar{\Pvec}^\al
+ \hbar^2 \, \Om^{\mu \al \bt} \, \Om_{\nu \al \bt}
- \hbar^2 \lt(\nabvec^\al \, \Om^\mu{}_{\nu \al} \rt) \,
\label{D^2-spin-1}
\ee
for the spin-1 Casimir scalar for linear momentum, where the second and third terms in
(\ref{D^2-spin-1}) are Hermitian, while the fourth term involving the gradient of the Ricci rotation coefficients
is anti-Hermitian.
In the absence of external fields, it is clear that (\ref{D^2-spin-1}) is a Lorentz invariant.

To compute the corresponding Casimir scalar for spin, it is useful to note that the Proca equation naturally generates
the constraint
\be
\bar{\Pvec}_\nu \, V^\nu_{(1)} & = & 0 \, .
\label{P.V=0}
\ee
Upon substituting
\be
\bar{\Pvec}_\nu \, \bar{\Pvec}^\mu & = &
\bar{\Pvec}^\mu \, \bar{\Pvec}_\nu  + i \hbar \lt[\varepsilon^{0j\mu}{}_\nu \, \Rvec_j
+ Q^\mu{}_\nu - e \lt({}^F F^\mu{}_\nu + C^{\al \mu}{}_\nu \, \Avec_\al\rt) \rt] \,
\label{P-invert}
\ee
into $\Wvec^{(1)} \cdot \Wvec^{(1)}$ and using (\ref{P.V=0}), it is shown that
\be
\lt[\Wvec^\al_{(1)} \, \Wvec_\al^{(1)}\rt]^\mu{}_\nu
& = & - 1(1 + 1) \lt\{m_0^2 \, \dl^\mu{}_\nu  
- i \hbar \lt[\varepsilon^{0j\mu}{}_\nu \, \Rvec_j + Q^\mu{}_\nu - e \lt({}^F F^\mu{}_\nu + C^{\al \mu}{}_\nu \, \Avec_\al\rt) \rt] \rt\}
\nn
&  &{} + 3 i \hbar \, \Om_\nu{}^{[\mu \al]} \, \bar{\Pvec}_\al
+ 2 i \hbar \lt[\dl^\mu{}_{[\nu} \, \Om_{\al] \sg}{}^\sg
- \eta^{\mu \sg} \Om_{\al (\sg \nu)}\rt]\bar{\Pvec}^\al
\nn
&  &{}
- 2 \hbar^2 \lt(\Om^\sg{}_{\al \sg} \Om_\nu{}^{[\mu \al]} + \Om^{\mu[\al \bt]} \, \Om_{\nu [\al \bt]} \rt)
- 4 \hbar^2 (\nabvec_\al \, \Om_\nu{}^{[\mu \al]}) \, ,
\label{W^2-spin-1}
\ee
which is a Hermitian operator, except for the gradient term in (\ref{W^2-spin-1}).
In contrast to (\ref{D^2-spin-1}), the spin-1 Casimir scalar for spin is {\em not}
a Lorentz invariant for general motion in the absence of external fields, but
rather takes the form
\be
\lt[\Wvec^\al_{(1)} \, \Wvec_\al^{(1)}\rt]^\mu{}_\nu
& = & - 1(1 + 1) \lt(m_0^2 \, \dl^\mu{}_\nu - i \hbar \, \varepsilon^{0j\mu}{}_\nu \, \Rvec_j \rt) \, ,
\label{W^2-spin-1-free-particle}
\ee
where the second term due to the non-inertial dipole operator $\Rvec$ can be interpreted as an effective mass contribution.
Issues pertaining to this interpretation are considered in greater detail in the next section of this paper.

As for the orthogonality relationship between $\Wvec^{(1)}$ and $\Dvec^{(1)}$, it is evident that
this condition is violated in the presence of gravitational and electromagnetic fields, since
%
\be
\lt[\Wvec^\mu_{(1)} \, \Dvec_\mu^{(1)}\rt]^\lm{}_\rho
& = & -{i \hbar \over 2} \lt\{\lt[\Sgvec^{0j}_{(1)} \, \Rvec_j\rt]^\lm{}_\rho
+ i \, \varepsilon^{\al \bt \lm}{}_\rho \lt[Q_{\al \bt} - e \lt({}^F F_{\al \bt} + C^\mu{}_{\al \bt} \, \Avec_\mu\rt) \rt]\rt\}
\nn
&  &{} + 2 \hbar \, \varepsilon^{\lm \mu \al \bt} \, \Om_{\rho \al (\bt} \, \bar{\Pvec}_{\mu)}
+ i \hbar^2 \, \varepsilon^{\lm \mu \al \bt} \lt[\Om_{\sg \al \bt} \, \Om^\sg{}_{\rho \mu}
+ \lt(\nabvec_\al \Om_{\bt \rho \mu} \rt) \rt] \, ,
\label{W.D-spin-1}
\nl
\nn
\lt[\Dvec^\mu_{(1)} \, \Wvec_\mu^{(1)}\rt]^\lm{}_\rho
& = & {i \hbar \over 2} \lt\{\lt[\Sgvec^{0j}_{(1)} \, \Rvec_j\rt]^\lm{}_\rho
+ i \, \varepsilon^{\al \bt \lm}{}_\rho \lt[Q_{\al \bt} - e \lt({}^F F_{\al \bt} + C^\mu{}_{\al \bt} \, \Avec_\mu\rt) \rt]\rt\}
\nn
&  &{} - \hbar \, \varepsilon^{\mu \sg \al \bt} \lt[\eta_{\sg \rho} \, \Om^\lm{}_{\al \mu} \, \bar{\Pvec}_\bt
+ \dl^\lm{}_\sg \, \Om_{\rho \al \bt} \, \bar{\Pvec}_\mu \rt]
\nn
&  &{}- i \hbar^2 \, \varepsilon^{\mu \sg \al \bt} \lt[\Om^\lm{}_{\sg \mu} \, \Om_{\rho \al \bt}
+ \lt(\nabvec_\mu \Om_{\rho \al \bt} \rt) \rt] \, .
\label{D.W-spin-1}
\ee
%
However, it is also evident that $\Wvec^{(1)}$ and $\Dvec^{(1)}$ anticommute in the form
$\lt\{\Wvec^\mu_{(1)} \, , \, \Dvec_\mu^{(1)}\rt\} = 0$ in the limit of vanishing external fields, since
\be
\lt[\Wvec^\mu_{(1)} \, \Dvec_\mu^{(1)}\rt]^\lm{}_\rho
& = & - \lt[\Dvec^\mu_{(1)} \, \Wvec_\mu^{(1)}\rt]^\lm{}_\rho
\ = \ -{i \hbar \over 2} \lt[\Sgvec^{0j}_{(1)} \, \Rvec_j\rt]^\lm{}_\rho \, .
\label{W.D+D.W=0-spin-1}
\ee
This property concurs with the same type of observation for spin-1/2 particles, as given by (\ref{W.D-spin-1/2}).

\subsection{Spin-3/2 Fields}

While it is possible to construct a spin-3/2 field in different ways \cite{Ramond},
arguably the most useful representation is that given by the Rarita-Schwinger formalism \cite{Rarita}.
For this representation, the spin-3/2 vector-spinor field $\Psi^\mu_{(3/2)}$ is subject to the local Lorentz transformation
\be
\Psi'^\mu_{(3/2)} & = & \lt(\dl^\mu{}_\nu
- {i \over 2} \, \om^{\al \bt} \lt[\Sgvec_{\al \bt}^{(3/2)}\rt]^\mu{}_\nu \rt) \Psi^\nu_{(3/2)} \, ,
\label{spin-3/2}
\ee
where the covariant spin operator $\lt[\Sgvec_{\al \bt}^{(3/2)}\rt]^\mu{}_\nu$ is
\be
\lt[\Sgvec_{\al \bt}^{(3/2)}\rt]^\mu{}_\nu & = & 2i \, \dl^\mu{}_{[\al} \eta_{\bt] \nu}
+ {1 \over 2} \sg_{\al \bt} \, \dl^\mu{}_\nu \, ,
\label{om-spin-3/2}
\ee
expressed as the sum of spin-1 interactions on the spin-3/2 field's vector components
and spin-1/2 interactions on the spinor components of $\Psi^\mu_{(3/2)}$.
Application of (\ref{D-general}) with (\ref{om-spin-3/2}) results in
\be
\lt[\Dvec_\mu^{(3/2)}\rt]^\lm{}_\rho & = & \dl^\lm{}_\rho \, \bar{\Pvec}_\mu
+ {\hbar \over 2} \lt[\Sgvec^{\al \bt}_{(3/2)}\rt]^\lm{}_\rho \, \Om_{\al \bt \mu}
\ = \ \dl^\lm{}_\rho \, \Dvec_\mu^{(1/2)} + i \hbar \, \Om^\lm{}_{\rho \mu} \,
\label{D-spin-3/2}
\ee
for the spin-3/2 covariant momentum operator.
The corresponding expression for the spin-3/2 Pauli-Lubanski vector is then
\be
\lt[\Wvec^\mu_{(3/2)} \rt]^\lm{}_\rho & = & - {1 \over 2} \, \varepsilon^{\mu \al \bt \gm}
\, \lt[\Sgvec_{\al \bt}^{(3/2)}\rt]^\lm{}_\sg \, \lt[\Dvec_\gm^{(3/2)}\rt]^\sg{}_\rho
\nn
& = & \lt[\Wvec^\mu_{(1/2)} \rt] \dl^\lm{}_\rho + \lt[\Wvec^\mu_{(1)} \rt]^\lm{}_\rho
\nn
& = &
\lt[- {1 \over 4} \, \varepsilon^\mu{}_{\al \bt \gm} \, \sg^{\al \bt}
\, \Dvec^\gm_{(1/2)} \rt] \dl^\lm{}_\rho \,
- i \, \varepsilon^{\mu \lm}{}_{\rho \gm} \, \bar{\Pvec}^\gm
- \hbar \, \varepsilon^{\mu \lm \al \bt} \, \Om_{\rho \al \bt} \,
\label{W-spin-3/2}
\ee
from using (\ref{W-general}), (\ref{om-spin-3/2}), and (\ref{D-spin-3/2}).

Straightforward application of (\ref{D-spin-3/2}) to compute $\Dvec^{(3/2)} \cdot \Dvec^{(3/2)}$ leads to
\be
\lt[\Dvec^\al_{(3/2)} \, \Dvec_\al^{(3/2)} \rt]^\mu{}_\nu & = &
\lt[\Dvec^\al_{(1/2)} \, \Dvec_\al^{(1/2)}\rt] \dl^\mu{}_\nu
+ \lt[\Lm^{(3/2)} \rt]^\mu{}_\nu \,
\label{D^2-spin-3/2}
\ee
for the spin-3/2 Casimir scalar for linear momentum, where
\be
\lt[\Lm^{(3/2)} \rt]^\mu{}_\nu & = & \lt[\Dvec^\al_{(1)} \, \Dvec_\al^{(1)} \rt]^\mu{}_\nu
- m_0^2 \, \dl^\mu{}_\nu - 2 i \hbar^2 \, \Om^\mu{}_{\nu \al}
\lt(\gm^5 \, \bar{\Gmvec}^\al_{\rm (C)} - i \, \bar{\Gmvec}^\al_{\rm (S)} \rt)
\nn
& = & 2i \hbar \, \Om^\mu{}_{\nu \al} \, \bar{\Pvec}^\al - 2 i \hbar^2 \, \Om^\mu{}_{\nu \al}
\lt(\gm^5 \, \bar{\Gmvec}^\al_{\rm (C)} - i \, \bar{\Gmvec}^\al_{\rm (S)} \rt) + \hbar^2 \, \Om^{\mu \al \bt} \, \Om_{\nu \al \bt}
\nn
&  &{} - \hbar^2 \lt(\nabvec^\al \, \Om^\mu{}_{\nu \al} \rt)
\, .
\label{Lm-spin-3/2}
\ee
It is clear from (\ref{D^2-spin-3/2}) and (\ref{Lm-spin-3/2}) that $\Dvec^{(3/2)} \cdot \Dvec^{(3/2)}$
is expressed as the sum of spin-1/2 and spin-1 contributions separately, with cross terms dependent
on the gravitational field.
Except for the gradient-dependent term in (\ref{Lm-spin-3/2}), $\Dvec^{(3/2)} \cdot \Dvec^{(3/2)}$ is a Hermitian operator.
Furthermore, (\ref{D^2-spin-3/2}) is the Lorentz invariant
$\lt[\Dvec^\al_{(3/2)} \, \Dvec_\al^{(3/2)} \rt]^\mu{}_\nu = m_0^2 \, \dl^\mu{}_\nu$ in the limit of vanishing external fields.

To determine the spin-3/2 Casimir scalar for spin, it is necessary to make use of the known constraints
for Rarita-Schwinger fields \cite{Rarita}, which are
\be
\bar{\Pvec}_\nu \, \Psi^\nu & = & 0 \,
\label{P.Psi=0}
\ee
and
\be
\gm_\nu \, \Psi^\nu & = & 0 \, .
\label{gm.Psi=0}
\ee
With (\ref{P.Psi=0}) and (\ref{gm.Psi=0}), it can be shown that $\Wvec^{(3/2)} \cdot \Wvec^{(3/2)}$
can be evaluated, leading to
\be
\lt[\Wvec^\al_{(3/2)} \, \Wvec_\al^{(3/2)}\rt]^\mu{}_\nu & = &
\lt[\Wvec^\al_{(1/2)} \, \Wvec_\al^{(1/2)}\rt] \dl^\mu{}_\nu
+ \lt[\Wvec^\al_{(1)} \, \Wvec_\al^{(1)}\rt]^\mu{}_\nu
\nn
&  &{} - \lt\{m_0^2 \, \dl^\mu{}_\nu - i \hbar \lt[\varepsilon^{0j\mu}{}_\nu \, \Rvec_j
+ Q^\mu{}_\nu - e \lt({}^F F^\mu{}_\nu + C^{\al \mu}{}_\nu \, \Avec_\al\rt) \rt] \rt\}
\nn
&  &{} + \hbar\lt(\dl^\mu{}_\nu \, \eta^{\sg \al} - i \, \sg^{\mu \al} \, \dl^\sg{}_\nu \rt)
\lt(\gm^5 \, \bar{\Gmvec}^{\rm (C)}_\sg - i \, \bar{\Gmvec}^{\rm (S)}_\sg\rt) \bar{\Pvec}_\al
\nn
&  &{} - \hbar \, \sg^{\al \bt} \, \Om_{\nu \al \bt} \, \Dvec^\mu_{(1/2)}
- 2 \hbar \, \sg^{\mu \al} \, \Om_{\nu [\al \gm]} \, \Dvec^\gm_{(1/2)}
\nn
&  &{} - {i \hbar^2 \over 2} \lt[\sg^{\al \bt} \, \lt(\nabvec^\mu \, \Om_{\nu \al \bt}\rt)
+ 2 \, \sg^{\mu \al} \lt(\nabvec^\gm \, \Om_{\nu [\al \gm]}\rt) \rt]
\nn
&  &{} + {i \hbar^2 \over 2} \lt[\dl^\mu{}_\nu \, \nabvec_\al \lt(\gm^5 \, \bar{\Gmvec}_{\rm (C)}^\al - i \, \bar{\Gmvec}_{\rm (S)}^\al\rt)
- \nabvec_\nu \lt(\gm^5 \, \bar{\Gmvec}_{\rm (C)}^\mu - i \, \bar{\Gmvec}_{\rm (S)}^\mu\rt) \rt]
\nn
&  &{}
+ {\hbar^2 \over 2} \, \sg^{\mu \al} \, \nabvec_\al \lt(\gm^5 \, \bar{\Gmvec}^{\rm (C)}_\nu - i \, \bar{\Gmvec}^{\rm (S)}_\nu\rt) \,
\label{W^2-spin-3/2}
\ee
in the presence of gravitational and electromagnetic fields.
A closer examination of (\ref{W^2-spin-3/2}) in the limit of vanishing external fields reveals that
\be
\lt[\Wvec^\al_{(3/2)} \, \Wvec_\al^{(3/2)}\rt]^\mu{}_\nu & = &
\lt[-{3 \over 2} \lt({3 \over 2} + 1\rt) \, m_0^2 + {\hbar \over 2} \, \lt(\sgvec \cdot \Rvec\rt) \rt] \dl^\mu{}_\nu
+ 3 i \hbar \, \varepsilon^{0j\mu}{}_\nu \, \Rvec_j \, ,
\label{W^2-spin-3/2-free-particle}
\ee
where $\Rvec$ appears in terms of both the spin-1/2 and spin-1 interactions that comprise the spin-3/2 field,
which generate effective mass contributions to (\ref{W^2-spin-3/2-free-particle}).

In a similar fashion as shown for spin-1/2 and spin-1 fields, the orthogonality relations between $\Wvec^{(3/2)}$
and $\Dvec^{(3/2)}$ are not satisfied in the presence of gravitational and electromagnetic fields, as shown by
%
\be
\lt[\Wvec^\mu_{(3/2)} \, \Dvec_\mu^{(3/2)} \rt]^\lm{}_\rho & = &
\lt[\Wvec^\mu_{(1/2)} \, \Dvec_\mu^{(1/2)} \rt] \dl^\lm{}_\rho + \lt[\Wvec^\mu_{(1)} \, \Dvec_\mu^{(1/2)} \rt]^\lm{}_\rho
\nn
&  &{} + i \hbar \lt[\Wvec^\mu_{(3/2)}\rt]^\lm{}_\sg \, \Om^\sg{}_{\rho \mu} \, ,
\label{W.D-spin-3/2}
\nl
\nn
\lt[\Dvec^\mu_{(3/2)} \, \Wvec_\mu^{(3/2)} \rt]^\lm{}_\rho & = &
\lt[\Dvec^\mu_{(1/2)} \, \Wvec_\mu^{(1/2)} \rt] \dl^\lm{}_\rho + \lt[\Dvec^\mu_{(1/2)} \, \Wvec_\mu^{(1)} \rt]^\lm{}_\rho
\nn
&  &{} + i \hbar \, \Om^\lm{}_{\sg \mu} \lt[\Wvec^\mu_{(3/2)}\rt]^\sg{}_\rho  \, ,
\label{D.W-spin-3/2}
\ee
%
where
%
\be
\lt[\Wvec^\mu_{(1)} \, \Dvec_\mu^{(1/2)} \rt]^\lm{}_\rho & = &
{\hbar \over 2} \lt\{- i \lt[\Sgvec^{0j}_{(1)} \, \Rvec_j\rt]^\lm{}_\rho
+ \varepsilon^{\al \bt \lm}{}_\rho \lt[Q_{\al \bt} - e \lt({}^F F_{\al \bt} + C^\mu{}_{\al \bt} \, \Avec_\mu\rt) \rt]\rt\}
\nn
&  &{} - \hbar \, \varepsilon^{\mu \lm \al \bt} \, \Om_{\rho \al \bt} \, \bar{\Pvec}_\mu
+ i \hbar \, \varepsilon^{\al \bt \lm}{}_\rho \lt(\gm^5 \, \bar{\Gmvec}^{(\rm C)}_\al - i \, \bar{\Gmvec}^{(\rm S)}_\al\rt)
\bar{\Pvec}_\bt
\nn
&  &{} + \hbar^2 \, \varepsilon^{\mu \lm \al \bt} \,
\Om_{\rho \al \bt} \lt(\gm^5 \, \bar{\Gmvec}^{(\rm C)}_\mu - i \, \bar{\Gmvec}^{(\rm S)}_\mu\rt)
\nn
&  &{}
- \hbar^2 \, \varepsilon^{\mu \lm}{}_{\rho \gm} \, \nabvec^\gm
\lt(\gm^5 \, \bar{\Gmvec}^{(\rm C)}_\mu - i \, \bar{\Gmvec}^{(\rm S)}_\mu\rt) \, ,
\label{(W-1).(D-1/2)-spin-3/2}
\nl
\nn
\lt[\Dvec^\mu_{(1/2)} \, \Wvec_\mu^{(1)} \rt]^\lm{}_\rho & = &
- {\hbar \over 2} \lt\{- i \lt[\Sgvec^{0j}_{(1)} \, \Rvec_j\rt]^\lm{}_\rho
+ \varepsilon^{\al \bt \lm}{}_\rho \lt[Q_{\al \bt} - e \lt({}^F F_{\al \bt} + C^\mu{}_{\al \bt} \, \Avec_\mu\rt) \rt]\rt\}
\nn
&  &{} - \hbar \, \varepsilon^{\mu \lm \al \bt} \, \Om_{\rho \al \bt} \, \bar{\Pvec}_\mu
+ i \hbar \, \varepsilon^{\al \bt \lm}{}_\rho \lt(\gm^5 \, \bar{\Gmvec}^{(\rm C)}_\al - i \, \bar{\Gmvec}^{(\rm S)}_\al\rt)
\bar{\Pvec}_\bt
\nn
&  &{} + \hbar^2 \, \varepsilon^{\mu \lm \al \bt} \,
\Om_{\rho \al \bt} \lt(\gm^5 \, \bar{\Gmvec}^{(\rm C)}_\mu - i \, \bar{\Gmvec}^{(\rm S)}_\mu\rt)
- i \hbar^2 \, \varepsilon^{\mu \lm \al \bt} \lt(\nabvec_\mu \Om_{\rho \al \bt} \rt) \, .
\label{(D-1/2).(W-1)-spin-3/2}
\ee
%
It is also true for spin-3/2 fields that $\Wvec^{(3/2)}$ and $\Dvec^{(3/2)}$
anticommute according to $\lt\{\Wvec^\mu_{(3/2)} \, , \, \Dvec_\mu^{(3/2)}\rt\} = 0$
in the limit of vanishing external fields, since
\be
\lt[\Wvec^\mu_{(3/2)} \, \Dvec_\mu^{(3/2)}\rt]^\lm{}_\rho
& = & - \lt[\Dvec^\mu_{(3/2)} \, \Wvec_\mu^{(3/2)}\rt]^\lm{}_\rho
\nn
& = & - \lt\{{\hbar \over 2} \lt(\alvec \cdot \Rvec\rt) \dl^\lm{}_\rho
+ {i \hbar \over 2} \lt[\Sgvec^{0j}_{(1)} \, \Rvec_j\rt]^\lm{}_\rho \rt\} \, .
\label{W.D+D.W=0-spin-3/2}
\ee

\subsection{Spin-2 Fields}

For a spin-2 field described by $T^{\mu\nu}_{(2)}$ as the composite of two spin-1 vector fields,
the local Lorentz transformation is given by
\be
T'^{\mu \nu}_{(2)} & = & \lt(\dl^\mu{}_\al \, \dl^\nu{}_\bt
- {i \over 2} \, \om^{\sg \lm} \lt[\Sgvec_{\sg \lm}^{(2)}\rt]^{\mu\nu}{}_{\al\bt} \rt)
T^{\al\bt}_{(2)} \, ,
\label{spin-2}
\ee
where the covariant spin operator $\lt[\Sgvec_{\sg \lm}^{(2)}\rt]^{\mu\nu}{}_{\al\bt}$ is
\be
\lt[\Sgvec_{\sg \lm}^{(2)}\rt]^{\mu\nu}{}_{\al\bt}
& = & 2i \lt(\dl^\mu{}_{[\sg} \eta_{\lm] \al} \, \dl^\nu{}_\bt
+ \dl^\nu{}_{[\sg} \eta_{\lm] \bt} \, \dl^\mu{}_\al \rt) \, .
\label{om-spin-2}
\ee
Following the same pattern as before, the spin-2 covariant momentum operator takes the form
\be
\lt[\Dvec_\mu^{(2)}\rt]^{\lm \xi}{}_{\rho \eta}
& = & \dl^\lm{}_\rho \, \dl^\xi{}_\eta \, \bar{\Pvec}_\mu
+ {\hbar \over 2} \lt[\Sgvec^{\al \bt}_{(2)}\rt]^{\lm\xi}{}_{\rho\eta} \, \Om_{\al \bt \mu}
\nn
& = & \dl^\lm{}_\rho \, \dl^\xi{}_\eta \, \bar{\Pvec}_\mu
+ i \hbar \lt(\dl^\lm{}_\rho \, \Om^\xi{}_{\eta \mu}
+ \dl^\xi{}_\eta \, \Om^\lm{}_{\rho \mu}  \rt)
\nn
& = & \dl^\lm{}_\rho \lt[\Dvec_\mu^{(1)} \rt]^\xi{}_\eta + \dl^\xi{}_\eta \lt[\Dvec_\mu^{(1)} \rt]^\lm{}_\rho
- \dl^\lm{}_\rho \, \dl^\xi{}_\eta \, \bar{\Pvec}_\mu
\, ,
\label{D-spin-2}
\ee
while the corresponding spin-2 Pauli-Lubanski vector is
\be
\lt[\Wvec^\mu_{(2)} \rt]^{\lm \xi}{}_{\rho {\eta}}
& = & - {1 \over 2} \, \varepsilon^{\mu \al \bt \gm}
\, \lt[\Sgvec_{\al \bt}^{(2)}\rt]^{\lm \xi}{}_{\rho' {\eta}'} \,
\lt[\Dvec_\gm^{(2)}\rt]^{\rho' {\eta}'}{}_{\rho {\eta}}
\nn
& = & \dl^\lm{}_\rho \lt[\Wvec^\mu_{(1)} \rt]^\xi{}_{{\eta}}
+ \lt[\Wvec^\mu_{(1)} \rt]^\lm{}_\rho \, \dl^\xi{}_{{\eta}}
+ \hbar \lt(\lt[\Omvec^\mu \rt]^{\lm \xi}{}_{\rho \eta}
+ \lt[\Omvec^\mu \rt]^{\xi \lm}{}_{\eta \rho} \rt) \, ,
\label{W-spin-2}
\ee
where $\lt[\Omvec^\mu \rt]^{\lm \xi}{}_{\rho \eta} \equiv \varepsilon^{\mu\al\lm}{}_\rho \, \Om^\xi{}_{\eta\al}$.

Computation of the spin-2 Casimir scalar for linear momentum is then given by
\be
\lt[\Dvec^\mu_{(2)} \, \Dvec_\mu^{(2)} \rt]^{\lm \xi}{}_{\rho \eta}
& = & \dl^\lm{}_\rho \lt\{\lt[\Dvec^\mu_{(1)} \, \Dvec_\mu^{(1)} \rt]^\xi{}_\eta
- 2 \, \lt[\Dvec^\mu_{(1)}\rt]^\xi{}_\eta \, \bar{\Pvec}_\mu + {1 \over 2} \, m_0^2 \, \dl^\xi{}_\eta \rt\}
\nn
&  &{} + \lt[\Dvec^\mu_{(1)}\rt]^\lm{}_\rho \lt[\Dvec_\mu^{(1)}\rt]^\xi{}_\eta
+ \hbar^2 \, \dl^\lm{}_\rho \lt(\nabvec^\mu \, \Om^\xi{}_{\eta \mu}\rt)
\nn
&  &{} + \lt[\lt(\lm , \, \rho\rt) \leftrightarrow \lt(\xi , \, \eta\rt)\rt]
\nn
& = & \dl^\lm{}_\rho \lt\{\lt[\Dvec^\mu_{(1)} \, \Dvec_\mu^{(1)} \rt]^\xi{}_\eta
- {1 \over 2} \, m_0^2 \, \dl^\xi{}_\eta + \hbar^2 \, \lt(\nabvec^\mu \, \Om^\xi{}_{\eta \mu}\rt) \rt\}
 - \hbar^2 \, \eta^{\lm \mu} \, \Om^\lm{}_{\rho \lm} \, \Om^\xi{}_{\eta \mu}
\nn
&  &{} + \lt[\lt(\lm , \, \rho\rt) \leftrightarrow \lt(\xi , \, \eta\rt)\rt]
\, ,
\label{D^2-spin-2}
\ee
where $\lt[\lt(\lm , \, \rho\rt) \leftrightarrow \lt(\xi , \, \eta\rt)\rt]$ denotes the addition of
identical terms with the interchange of index pairs as given.
This expression for $\Dvec^{(2)} \cdot \Dvec^{(2)}$ also reduces to the Lorentz invariant
$\lt[\Dvec^\mu_{(2)} \, \Dvec_\mu^{(2)} \rt]^{\lm \xi}{}_{\rho \eta} = m_0^2 \, \dl^\lm{}_\rho \, \dl^\xi{}_\eta$
in the absence of external fields.
As for the corresponding spin-2 Casimir scalar for spin, this is given by
\be
\lt[\Wvec^\mu_{(2)} \, \Wvec_\mu^{(2)} \rt]^{\lm \xi}{}_{\rho \eta}
& = & \dl^\lm{}_\rho \, \lt[\Wvec^\mu_{(1)} \, \Wvec_\mu^{(1)} \rt]^\xi{}_\eta
+ \lt[\Wvec^\mu_{(1)}\rt]^\lm{}_\rho \, \lt[\Wvec_\mu^{(1)}\rt]^\xi{}_\eta
\nn
&  &{} + \hbar \lt[\Wvec^\mu_{(1)}\rt]^\lm{}_\al
\lt(\lt[\Omvec_\mu\rt]^{\al \xi}{}_{\rho \eta} + \lt[\Omvec_\mu\rt]^{\xi \al}{}_{\eta \rho}\rt)
\nn
&  &{} + \hbar \lt(\lt[\Omvec^\mu\rt]^{\lm \xi}{}_{\rho \al} + \lt[\Omvec^\mu\rt]^{\xi \lm}{}_{\al \rho}\rt)
\lt[\Wvec_\mu^{(1)}\rt]^\al{}_\eta
\nn
&  &{} + \hbar^2 \lt(\lt[\Omvec^\mu \, \Omvec_\mu \rt]^{\lm \xi}{}_{\rho \eta}
+ \lt[\Omvec^\mu\rt]^{\lm \xi}{}_{\al \bt} \, \lt[\Omvec_\mu\rt]^{\bt \al}{}_{\eta \rho} \rt)
\nn
&  &{} + \lt[\lt(\lm , \, \rho\rt) \leftrightarrow \lt(\xi , \, \eta\rt)\rt]
\, ,
\label{W^2-spin-2}
\ee
where the constraint equation
\be
\bar{\Pvec}_\nu \, T^{\mu \nu} & = & 0 \,
\label{P.T=0}
\ee
projecting out the longitudinal modes is incorporated.
In the absence of external fields, it is shown that (\ref{W^2-spin-2}) reduces to
\be
\lt[\Wvec^\mu_{(2)} \, \Wvec_\mu^{(2)} \rt]^{\lm \xi}{}_{\rho \eta}
& = & - \lt(2 \, \dl^\lm{}_\rho \, \dl^\xi{}_\al + \dl^\xi{}_\rho  \, \dl^\lm{}_\al \rt)
\lt(m_0^2 \, \dl^\al{}_\eta - i \hbar \, \varepsilon^{0j\al}{}_\eta \, \Rvec_j \rt)
\nn
&  &{} + \lt(m_0^2 \, \eta^{\lm \xi} - \Pvec^\xi \, \Pvec^\lm \rt) \eta_{\rho \eta}
+ \lt[\lt(\lm , \, \rho\rt) \leftrightarrow \lt(\xi , \, \eta\rt)\rt]
\, .
\label{W^2-spin-2-free-particle}
\ee
So far, no symmetries are assumed for $T^{\mu\nu}_{(2)}$ in the derivations leading to (\ref{W^2-spin-2-free-particle}).
However, when considering a symmetric and traceless spin-2 tensor field, commonly accepted defining properties for the graviton
$h^{\mu \nu}_{(2)}$ \cite{Ramond,Feynman}, it follows naturally that
\be
\lt[\Wvec^\mu_{(2)} \, \Wvec_\mu^{(2)} \rt]^{\lm \xi}{}_{\rho \eta} 
& = & -2 \lt(2 + 1 \rt)
\lt[m_0^2 \, \dl^\lm{}_\rho \, \dl^\xi{}_\eta - 2 i \hbar \, \varepsilon^{0j\lm}{}_{(\rho} \, \dl^\xi{}_{\eta)} \, \Rvec_j\rt]
\, ,
\label{W^2-spin-2-massive-graviton}
\ee
where the non-inertial dipole operator $\Rvec$ again appears as an effective mass contribution in (\ref{W^2-spin-2-massive-graviton}).

Finally, following the same pattern as found in lower-order spins, it is shown that the orthogonality relationship between
$\Wvec^{(2)}$ and $\Dvec^{(2)}$ is no longer satisfied in the presence of gravitational and electromagnetic fields, since
%
\be
\lt[\Wvec^\mu_{(2)} \, \Dvec_\mu^{(2)} \rt]^{\lm \xi}{}_{\rho \eta} & = & \lt[\Wvec^\mu_{(1)}\rt]^\lm{}_\rho \lt[\Dvec_\mu^{(1)}\rt]^\xi{}_\eta
+ \lt[\Wvec^\mu_{(1)}\rt]^\xi{}_\eta \lt[\Dvec_\mu^{(1)}\rt]^\lm{}_\rho
\nn
&  &{} + \dl^\lm{}_\rho \lt[\Wvec^\mu_{(1)} \lt(\Dvec_\mu^{(1)} - \bar{\Pvec}_\mu\rt) \rt]^\xi{}_\eta
+ \lt[\Wvec^\mu_{(1)} \lt(\Dvec_\mu^{(1)} - \bar{\Pvec}_\mu\rt) \rt]^\lm{}_\rho \, \dl^\xi{}_\eta
\nn
&  &{} + \hbar \lt(\varepsilon^{\mu \gm \lm}{}_\rho \, \Om^\xi{}_{\sg \gm} + \varepsilon^{\mu \gm \xi}{}_\sg \, \Om^\lm{}_{\rho \gm} \rt)
\lt[\Dvec_\mu^{(1)}\rt]^\sg{}_\eta
\nn
&  &{} + i \hbar^2 \lt(\varepsilon^{\mu \gm \lm}{}_\sg \, \Om^\xi{}_{\eta \gm} + \varepsilon^{\mu \gm \xi}{}_\eta \, \Om^\lm{}_{\sg \gm} \rt)
\Om^\sg{}_{\rho \mu} \, ,
\label{W.D-spin-2}
\nl
\nn
\lt[\Dvec^\mu_{(2)} \, \Wvec_\mu^{(2)} \rt]^{\lm \xi}{}_{\rho \eta} & = & \lt[\Dvec^\mu_{(1)}\rt]^\lm{}_\rho \lt[\Wvec_\mu^{(1)}\rt]^\xi{}_\eta
+ \lt[\Dvec^\mu_{(1)}\rt]^\xi{}_\eta \lt[\Wvec_\mu^{(1)}\rt]^\lm{}_\rho
\nn
&  &{} + \dl^\lm{}_\rho \lt[\lt(\Dvec^\mu_{(1)} - \bar{\Pvec}^\mu\rt) \Wvec_\mu^{(1)}\rt]^\xi{}_\eta
+ \lt[\lt(\Dvec^\mu_{(1)} - \bar{\Pvec}^\mu\rt) \Wvec_\mu^{(1)}\rt]^\lm{}_\rho \, \dl^\xi{}_\eta
\nn
&  &{} + \hbar \lt(\varepsilon^{\mu \gm \lm}{}_\rho \, \Om^\sg{}_{\eta \gm} + \varepsilon^{\mu \gm \sg}{}_\eta \, \Om^\lm{}_{\rho \gm} \rt)
\lt[\Dvec_\mu^{(1)}\rt]^\xi{}_\sg
\nn
&  &{} + i \hbar^2 
\lt(\varepsilon^{\mu \gm \sg}{}_\rho \, \Om^\xi{}_{\eta \gm} + \varepsilon^{\mu \gm \xi}{}_\eta \, \Om^\sg{}_{\rho \gm} \rt)
\Om^\lm{}_{\sg \mu}
\nn
&  &{} + i \hbar^2
\lt[\varepsilon^{\mu \gm \lm}{}_\rho \lt(\nabvec_\mu \Om^\sg{}_{\eta \gm}\rt)
+ \varepsilon^{\mu \gm \sg}{}_\eta \lt(\nabvec_\mu \Om^\lm{}_{\rho \gm}\rt) \rt] \dl^\xi{}_\sg \, .
\label{D.W-spin-2}
\ee
%
Again, following the same pattern as shown in lower-order spins,
in the limit of vanishing external fields, $\Wvec^{(2)}$ and $\Dvec^{(2)}$ anticommute in the form
$\lt\{\Wvec^\mu_{(2)} \, , \, \Dvec_\mu^{(2)}\rt\}= 0$, since
\be
\lt[\Wvec^\mu_{(2)} \, \Dvec_\mu^{(2)} \rt]^{\lm \xi}{}_{\rho \eta} & = & -\lt[\Dvec^\mu_{(2)} \, \Wvec_\mu^{(2)} \rt]^{\lm \xi}{}_{\rho \eta}
\nn
& = & -{i\hbar \over 2} \lt\{\dl^\lm{}_\rho \lt[\Sgvec^{0j}_{(1)} \, \Rvec_j\rt]^\xi{}_\eta
+ \dl^\xi{}_\eta \lt[\Sgvec^{0j}_{(1)} \, \Rvec_j\rt]^\lm{}_\rho \rt\} \, .
\label{W.D+D.W=0-spin-2}
\ee

\section{Casimir Spin Scalar Operators in Diagonalized Form}
\label{Sec.Casimir.diagonlized-spin}

\subsection{General Expressions}

At this point, it is useful to determine the eigenvalues of $\Wvec^{(s)} \cdot \Wvec^{(s)}$
in the absence of external fields and explore their physical implications,
given that each of the expressions for spin-1/2 to spin-2 inclusive have
non-trivial effective mass contributions present due to $\Rvec$.
This is a straightforward exercise, where the ``(D)'' subscript for $\lt[\Wvec^{(s)} \cdot \Wvec^{(s)}\rt]_{(\rm D)}$
in the computations that follow indicates the diagonalized form of the Casimir spin scalar operators.

For spin-1/2 particles, the diagonalized form of (\ref{W^2-spin-1/2}) in the absence of external fields leads to
\be
\lt[\Wvec^\al_{(1/2)} \, \Wvec_\al^{(1/2)}\rt]_{(\rm D)} & = &
-{1 \over 2} \lt({1 \over 2} + 1\rt) \lt(m_0^2 \pm \kappa_{1/2} \, \hbar |\Rvec| \rt) \, ,
\label{W^2-spin-1/2-free-particle-diag}
\ee
where $\kappa_{1/2} = {2 \over 3}$, which indicates an effective mass-squared splitting of
${1 \over 2} \, \hbar |\Rvec|$ above and below the standard eigenvalue of
$-{1 \over 2} \lt({1 \over 2} + 1\rt) m_0^2$.
Similarly, when considering spin-3/2, the diagonalized form of (\ref{W^2-spin-3/2-free-particle}) results in
\be
\lt[\Wvec^\al_{(3/2)} \, \Wvec_\al^{(3/2)}\rt]^\mu_{(\rm D) \, \nu} & = &
-{3 \over 2} \lt({3 \over 2} + 1\rt) \lt(m_0^2 \pm \kappa_{3/2} \, \hbar |\Rvec| \rt) \dl^\mu{}_\nu \, ,
\label{W^2-spin-3/2-free-particle-diag}
\ee
where $\kappa_{3/2} = {7 \over 3}, \, {11 \over 3}$, leading to effective mass-squared splittings of
${35 \over 4} \, \hbar |\Rvec|$ and ${55 \over 4} \, \hbar |\Rvec|$,
respectively, above and below $-{3 \over 2} \lt({3 \over 2} + 1\rt) m_0^2$.

In contrast, the spin-1 and spin-2 particle Casimir scalars each generate an eigenvalue
which has no dependence on $\Rvec$.
That is, for spin-1 particles it is shown from diagonalizing (\ref{W^2-spin-1-free-particle}) that
\be
\lt[\Wvec^\al_{(1)} \, \Wvec_\al^{(1)}\rt]^\mu_{(\rm D) \, \nu} & = &
-1 \lt(1 + 1\rt) \lt(m_0^2 \pm \kappa_1 \, \hbar |\Rvec|\rt) \dl^\mu{}_\nu \, ,
\label{W^2-spin-1-free-particle-diag}
\ee
where $\kappa_1 = 0, \, 1$, which indicates an effective mass-squared splitting of $2 \hbar |\Rvec|$ above and below the standard eigenvalue of
$-1 \lt(1 + 1\rt) m_0^2$, while diagonalizing (\ref{W^2-spin-2-massive-graviton}) for a massive symmetric spin-2 field
and adding the total number of excitations possible leads to
\be
\lt[\Wvec^\al_{(2)} \, \Wvec_\al^{(2)} \rt]^{\lm \xi}_{(\rm D) \, \rho \eta} 
& = & -2 \lt(2 + 1 \rt)
\lt(m_0^2 \pm \kappa_2 \, \hbar \, |\Rvec| \rt) \dl^\lm{}_\rho \, \dl^\xi{}_\eta \, ,
\label{W^2-spin-2-free-particle-diag}
\ee
where $\kappa_2 = 0, \, 1, \, 2$, with a corresponding effective mass-squared splitting of $6 \hbar |\Rvec|$
and $12 \hbar |\Rvec|$ above and below $-2 \lt(2 + 1 \rt) m_0^2$, respectively.

This outcome of $2s + 1$ eigenvalues with integer units of $\kappa_s \, \hbar |\Rvec|$ scaled by $s(s + 1)$ is analogous to the
well-understood notion of spectral line splitting due to the Zeeman effect, where $\Rvec$ fulfills the role of a
static magnetic field in this context, and $\kappa_s$ is the ``magnetic quantum number'' corresponding to excitations
with respect to non-inertial motion.
While it remains to be seen whether this analogy holds true in all conceivable contexts, given the
${\hbar \over 2} \lt(\sgvec \cdot \Rvec\rt)$ interaction term derived in (\ref{W^2-spin-1/2}) for spin-1/2 particles,
it appears that this interpretation is well-founded.
It is also interesting to note that the non-inertial dipole operator has the capacity to distinguish between fermions and bosons
due to this type of interaction with the spin.
Though it is unclear whether this pattern remains satisfied for $s > 2$, the fact that the distinction
between fermions and bosons due to $\Rvec$-dependent interactions applies to the most relevant spin states
known is a useful feature that may have future applications for consideration.

\subsection{Physical Implications for Massive and Massless Particles}

\begin{figure*}
\psfrag{E+}[cl][][2.5][0]{\large $+\kappa_s \, \hbar |\Rvec|$}
\psfrag{E-}[cl][][2.5][0]{\large $-\kappa_s \, \hbar |\Rvec|$}
\psfrag{Ws}[cc][][2.5][0]{\large $\Wvec^{(s)}$}
\hspace{1cm}
\begin{minipage}[t]{0.3 \textwidth}
\centering
\subfigure[\hspace{0.2cm} Spacelike Rotation]{
\label{fig:spacelike}
\rotatebox{0}{\includegraphics[width = 6cm, height = 6cm, scale = 1]{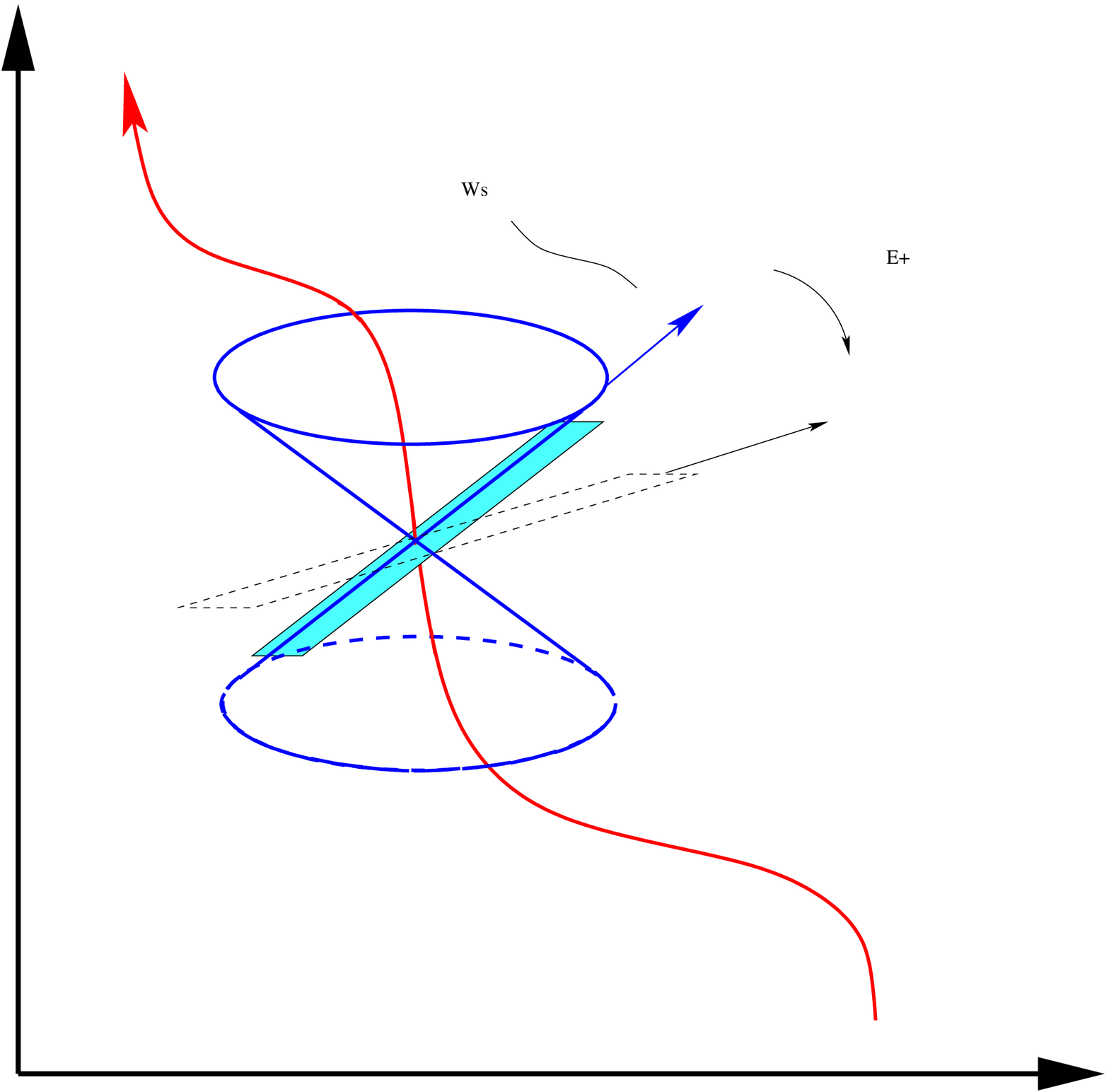}}}
\end{minipage}%
\hspace{3.0cm}
\begin{minipage}[t]{0.3 \textwidth}
\centering
\subfigure[\hspace{0.2cm} Timelike Rotation]{
\label{fig:timelike}
\rotatebox{0}{\includegraphics[width = 6cm, height = 6cm, scale = 1]{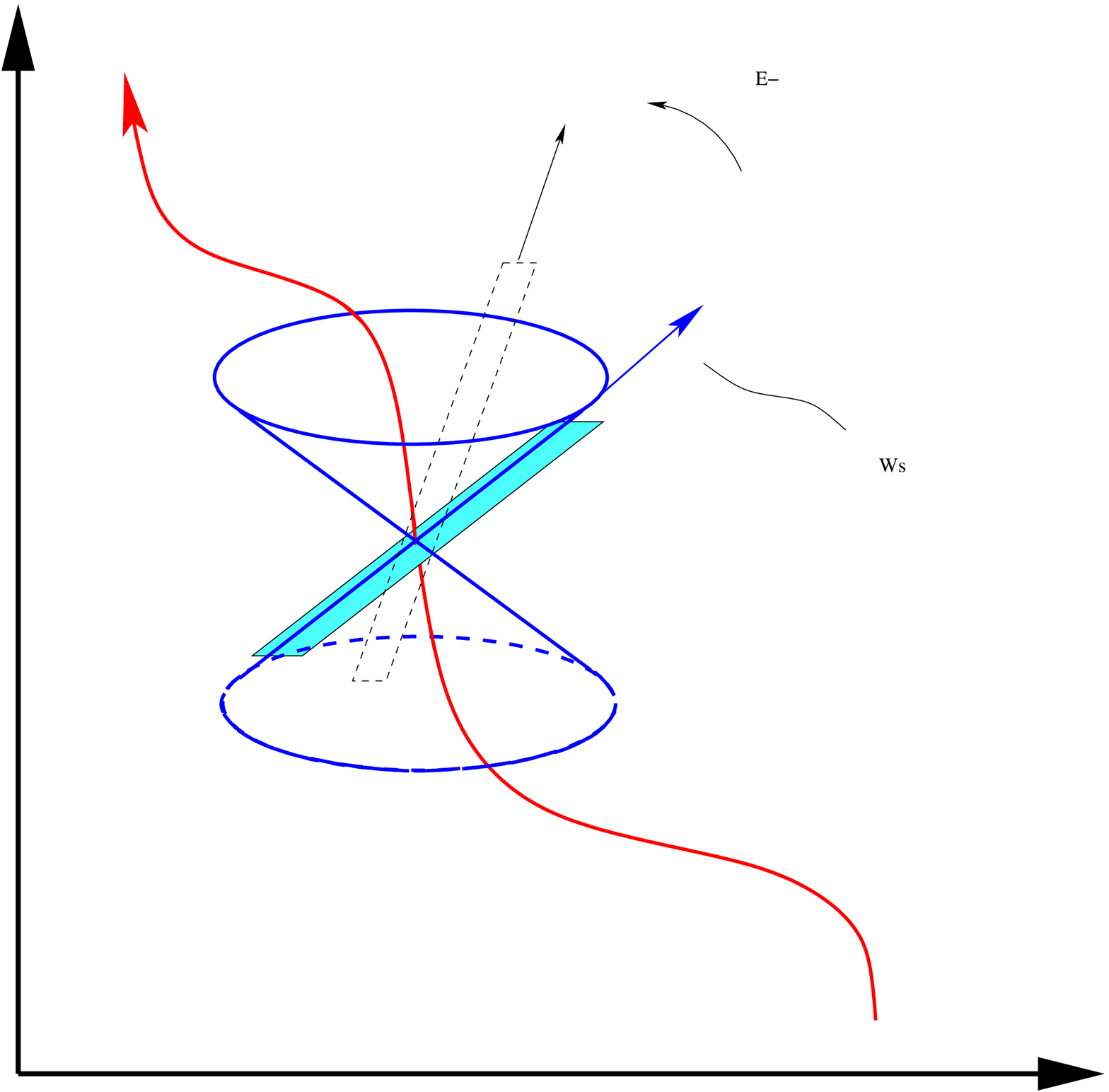}}}
\end{minipage}
\caption{\label{fig:light-cone} Rotation of the null plane containing $\Wvec^{(s)}$, as defined for massless particles,
due to non-inertial motion.
Fig.~\ref{fig:spacelike} transforms the null plane into a spacelike surface
for positive effective squared-mass $+\kappa_s \, \hbar |\Rvec|$ due to non-inertial motion,
while a negative effective squared-mass $-\kappa_s \, \hbar |\Rvec|$ leads to Fig.~\ref{fig:timelike}, a timelike surface.}
\end{figure*}
The eigenstates (\ref{W^2-spin-1/2-free-particle-diag})-(\ref{W^2-spin-2-free-particle-diag}) for spin-1/2 to spin-2
inclusive introduce some interesting physical consequences when applied to both massive
and massless spin-$s$ particles.
Consider for the moment the case of massive particles.
Assuming from the lower signs of (\ref{W^2-spin-1/2-free-particle-diag})-(\ref{W^2-spin-2-free-particle-diag}) the condition that
$m_0^2 - \kappa_s \, \hbar |\Rvec| > 0$, where $|\Rvec| \sim |\Pvec|/r$ and $\lm_0 = \hbar/m_0$,
it is straightforward to identify a critical speed $v_{\rm crit.} > v$, such that for $|\Pvec| = \gm \, m_0 v$
with Lorentz factor $\gm = 1/\sqrt{1 - v^2}$,
\be
v_{\rm crit.} & = & {1 \over \lt[1 + (\kappa_s \, \lm_0/r)^2 \rt]^{1/2}} \, .
\label{v-crit}
\ee
When $r \gg \kappa_s \, \lm_0$ to approximate free-particle inertial motion, $v_{\rm crit.} \rightarrow 1$ as expected.
However, when $r \sim \kappa_s \, \lm_0$, it follows that $v_{\rm crit.} \rightarrow 1/\sqrt{2}$.
The identification of a potential critical speed $v_{\rm crit.} < 1$ is interesting for at least two reasons.
First, it suggests the existence of a {\em maximal acceleration} $a_{\rm max.} \equiv 2/\lm_0$
originally proposed by Caianiello \cite{Caianiello,Papini} if it proves impossible to violate
the condition $v < v_{\rm crit.}$ for particle motion along a general worldline.
Second, the actual value for $v_{\rm crit.}$ coincides with the value determined by Chicone and Mashhoon
as the critical initial speed when classical matter propagating along the symmetry axis of Kerr black holes
becomes tidally accelerated to form high-energy astrophysical jets \cite{Chicone2}.
Though it is unclear whether this correspondence is nothing more than a coincidence, this value for
$v_{\rm crit.}$ does suggest the possibility of a deeper connection worthy of further investigation.

For strictly massless particles, $|\Pvec| = E$, where $E$ is the particle's energy.
It follows that the null surface corresponding to $\Wvec^{(s)} \cdot \Wvec^{(s)}$ becomes {\em spacelike} for $+ \kappa_s \, \hbar |\Rvec|$
and {\em timelike} for $- \kappa_s \, \hbar |\Rvec|$.
This amounts to a rotation of the $\Wvec^{(s)} \cdot \Wvec^{(s)}$ null surface away from the light cone,
as indicated by Figure~\ref{fig:light-cone}, to become either a surface of simultaneity or a casual surface for particles with
positive or negative energy, respectively.
This effect is only relevant for $\Wvec^{(s)} \cdot \Wvec^{(s)}$, since $\Dvec^{(s)} \cdot \Dvec^{(s)} = 0$
still remains a Lorentz invariant in the absence of external fields.
It is unclear what to make of a timelike $\Wvec^{(s)}$ for this scenario, or even if it makes physical sense
to contemplate its existence, since photons and gravitons are reasonably presumed to only carry positive energy.
This may be an issue worth considering in detail at a later date.

\section{Discussion}
\label{Sec.discussion}

Having now explored the so-called Casimir scalar properties for spin-1/2 to spin-2 particles inclusive
for general particle motion while in the presence of external gravitational and electromagnetic fields,
it is useful to consider some more general issues of relevance.
Given the difficulties noted from (\ref{Casimir-momentum-violation})-(\ref{[W,P]}) which clearly show
that the Casimir invariance properties of $\Dvec^{(s)} \cdot \Dvec^{(s)}$ and $\Wvec^{(s)} \cdot \Wvec^{(s)}$
generally break down for non-inertial motion, it is worthwhile to investigate the possibility of identifying
genuine Casimir invariants corresponding to linear momentum and spin angular momentum for general particle motion in space-time.
At least for the case of spin-1/2 particles \cite{Singh-Mobed}, it is possible to construct a modified Pauli-Lubanski vector
that takes into account the existence of $\Rvec$.
Besides satisfying Lorentz invariance, this new operator possesses some attractive properties in terms of naturally identifying
helicity states, for example, that equally apply to both massive and massless spin-1/2 particles.
It would be most interesting to know if such constructions can be determined for particles of higher-order spin,
and whether similar properties identified for the spin-1/2 case are replicated accordingly \cite{Singh-Mobed}.


One of the primary motivations of this paper is to consider how gravitational and electromagnetic fields, as
well as the non-inertial dipole operator, modify the properties of massless spin-1 and spin-2 particles,
corresponding to photon and gravitons, as well as spin-3/2 particles for gravitinos predicted to exist
in supersymmetry (SUSY).
That non-trivial contributions appear due to external fields is undeniable and may well provide useful insights into
how these mediating particles propagate in space-time.
For massive particles, it is reasonable to identify a world-tube of radius $\lm_0$ to describe
spatially-averaged quantum fluctuations surrounding a classical worldline where the propagating particle is most likely
to be found.
However, it is also known that an attempt to describe massless particles by assuming $m_0 \rightarrow 0$
within this framework leads to conceptual difficulties \cite{Goldhaber}, since the associated Compton wavelength
for the photon or graviton is now $\lm_0 \rightarrow \infty$.
This makes the world-tube treatment very difficult to conceptualize properly, since the photon or graviton must somehow
still resolve to propagate on null rays to preserve a light cone structure in the classical limit.

The well-known treatment of Fierz and Pauli \cite{Goldhaber,Fierz} assumes a massive spin-2 symmetric tensor for the
graviton propagating on a flat space-time background, which results in a modified set of Einstein field equations
due to an additional mass term that changes the geometric structure of the theory.
While the helicity states of $\pm 1$ for the massive graviton get removed as four-divergences in combination with
energy-momentum conservation, it happens that the zero-helicity state is coupled to the trace of the energy-momentum tensor
and still remains \cite{Goldhaber}.
Therefore, Einstein's general relativity is {\em not} recovered as $m_0 \rightarrow 0$, but behaves like a Brans-Dicke theory
of gravity.
With the external field contributions to the Casimir scalars considered in this paper, it may be possible
to overcome some of the limitations encountered by the Fierz-Pauli treatment, while also making use of the related approach
given by Feynman \cite{Feynman}.

The spin-3/2 gravitino described in terms of the Rarita-Schwinger construction is naturally a more speculative consideration
that exists within the SUSY formalism as the fermion superpartner with the graviton \cite{Steffen}.
It is widely regarded as a dark matter candidate to help account for the currently accepted value of the
known Universe's matter density.
The gravitino is classified as an extremely weakly interacting particle
that must be massless, but whose theoretical rest mass can exist from the eV to the TeV scale,
depending on the precise form of SUSY breaking \cite{Steffen}.
If such a particle truly exists, it is quite possible that non-inertial effects may contribute
significantly to understanding any mass determinations for relic gravitinos that may have been produced in the early Universe.

Finally, the approach taken in this paper assumes a smooth manifold and a simply connected topology for space-time.
This is certainly a reasonable assumption to make in the absence of any physical justification to suggest otherwise.
Of course, it is quite possible that space-time may be turbulent and have a highly non-trivial topological structure
for length scales of possible relevance for the purposes of this investigation.
The ``space-time foam'' concept envisioned by Wheeler \cite{Wheeler} may seriously disrupt the main features
of the statements given in this paper, though it is highly unclear when to anticipate the appearance of significant
space-time fluctuations, should the amplitude somehow be orders of magnitude larger than the Planck scale.
Nonetheless, there exist interesting approaches to this problem by suggesting searches for
quantum space-time fluctuations on cosmological scales \cite{Ng}, such as the presence of halos surrounding
distant quasars and other astrophysical sources.
It remains to be seen what sort of impact this outcome would likely have on the investigation presented here.

\section{Conclusion}
\label{Sec.conclusion}

This paper investigates the general properties of Casimir scalars for the Poincar\'{e}
group in the presence of external gravitational and electromagnetic fields, where the
group generators are expressed in terms of curvilinear co-ordinates to accommodate for
the general motion of elementary particles in space.
It is shown that under these conditions, the Casimir scalars for linear momentum and
spin angular momentum no longer satisfy the necessary conditions required
to denote them as ``Casimir invariants,'' since they no longer commute with all the
elements of the associated Lie algebra, according to (\ref{Casimir-momentum-violation})--(\ref{[W,P]}).
Explicit computations are performed for the specific cases of spin-1/2 to spin-2 particles inclusive,
where numerous interesting physical consequences emerge due to the presence of the non-inertial
dipole operator $\Rvec$, the most intriguing of which is the prediction of an effective
mass associated with the spin Casimir scalar.

While the external gravitational and electromagnetic contributions are given in general form,
it would be interesting to see how modifications of the Casimir scalars would appear
due to interactions with a specific metric background and/or charge distribution.
In particular, integration over quantum fluctuations about the elementary particle's classical worldline
would determine if there is an effective gravitational interaction that contributes to
the particle's propagation in space-time.
These and other issues outlined in this paper may be explored in the future.

\section{Acknowledgements}

The authors wish to thank Friedrich Hehl for identifying two references of interest for this paper \cite{Gronwald,Lemke},
and also Bahram Mashhoon for helpful comments about the paper.
They also acknowledge financial support from the University of Regina, Faculty of Science.

\appendix
\section{Derivation of the Momentum Commutator}
\label{Appendix-A}

Derivation (\ref{P-commutator}) requires use of (\ref{Pvec})--(\ref{Ovec}), which results in
\be
i \lt[\Pvec_{\hat{\al}} , \Pvec_{\hat{\bt}}\rt] & = & i \lt\{
(i \hbar)^2 \, \lt[\nabvec_{\hat{\al}} , \nabvec_{\hat{\bt}} \rt]
+ 2 i \hbar \, \nabvec_{[\hat{\al}} \, \Om_{\hat{\bt}]} \rt\}
\nn
& = & (i \hbar)^2 \lt\{i \lt[\nabvec_{\hat{\al}} , \nabvec_{\hat{\bt}} \rt]
\lt[1 +  \ln \lt(\lm^{\hat{1}}(u) \, \lm^{\hat{2}}(u) \, \lm^{\hat{3}}(u)\rt)^{1/2} \rt] \rt\} \, .
\label{A1}
\ee
From the orthonormal vierbein set $\lt\{e^\sg{}_{\hat{\al}} \rt\}$ to describe
$\nabvec_{\hat{\al}} = e^\sg{}_{\hat{\al}} \, \nabla_\sg$ in terms of Fermi normal co-ordinates,
it is shown that
\be
\lt[\nabvec_{\hat{\al}} , \nabvec_{\hat{\bt}} \rt] & = &
- 2 \lt(\nabla_\lm \, e^\sg{}_{[\hat{\al}} \rt) \, e^\lm{}_{\hat{\bt}]} \, \nabla_\sg \, ,
\label{A2}
\ee
where
\be
\lt(\nabla_\lm \, e^\sg{}_{[\hat{\al}} \rt) \, e^\lm{}_{\hat{\bt}]} \, \nabla_\sg  & = &
\bar{e}^\gm{}_{[\hat{\al}} \, \bar{e}^\lm{}_{\hat{\bt}]} \,
\lt[{\partial \over \partial X^\lm} \lt(\partial U^\sg \over \partial X^\gm\rt)\rt] {\partial \over \partial U^\sg}
\nn
&  &{} + \lt({\partial \over \partial U^\lm} \, \bar{e}^\gm{}_{[\hat{\al}} \rt) e^\lm{}_{\hat{\bt}]}
\lt({\partial U^\sg \over \partial X^\gm} \, {\partial \over \partial U^\sg}\rt) \, .
\label{A3}
\ee
By identifying
\be
{\partial \over \partial X^\gm} & = & {\partial U^\sg \over \partial X^\gm} \, {\partial \over \partial U^\sg}
\label{A4}
\ee
with
\be
\nabvec_{\hat{\gm}} & = & {1 \over \lm^{(\gm)}} \, {\partial \over \partial U^\gm} \, ,
\label{A5}
\ee
which is always possible by a suitable rotation of the orthonormal frame, it follows that
the first term in (\ref{A3}) can be described according to
\be
\lt[{\partial \over \partial X^\lm}
\lt(\partial U^\sg \over \partial X^\gm\rt)\rt] {\partial \over \partial U^\sg}
& \rightarrow & \lt[\nabvec_{\hat{\lm}} \, \lt(\lm^{(\gm)}\rt)^{-1}\rt] \, {\partial \over \partial U^\gm}
\nn
& = & - \lt[\nabvec_{\hat{\lm}} \, \ln \, \lm^{(\gm)}\rt] \nabvec_{\hat{\gm}} \, .
\label{A6}
\ee
Substitution of (\ref{A2}), (\ref{A3}), and (\ref{A6}) into (\ref{A1}) then leads to (\ref{P-commutator}),
where the second term of (\ref{A3}) corresponds to $\hbar \, C^{\hat{\mu}}{}_{\hat{\al}\hat{\bt}} \, \Pvec_{\hat{\mu}}$.

%
%

\end{document}